\pdfoutput=1

\documentclass[11pt,twoside,a4paper,cmspaper,final,collab]{cms-tdr}
\def\svnVersion{658ec61-D}\def\svnDate{2020/05/18}\def\cmsCernNoTag{CERN-EP-2020-011}\def\cmsCernDate{\today}\def\cmsMessage{"Published in the Journal of High Energy Physics as \href{http://dx.doi.org/10.1007/JHEP07(2020)125}{\doi{10.1007/JHEP07(2020)125}.}"}

\begin{document}\cmsNoteHeader{TOP-18-002}

\hyphenation{had-ron-i-za-tion}
\hyphenation{cal-or-i-me-ter}
\hyphenation{de-vices}
\RCS$HeadURL$
\RCS$Id$
\newlength\cmsFigWidth
\ifthenelse{\boolean{cms@external}}{\setlength\cmsFigWidth{0.49\textwidth}}{\setlength\cmsFigWidth{0.65\textwidth}}
\ifthenelse{\boolean{cms@external}}{\providecommand{\cmsLeft}{upper\xspace}}{\providecommand{\cmsLeft}{left\xspace}}
\ifthenelse{\boolean{cms@external}}{\providecommand{\cmsRight}{lower\xspace}}{\providecommand{\cmsRight}{right\xspace}}

\providecommand{\cmsTable}[1]{\resizebox{\textwidth}{!}{#1}}
\newlength\cmsTabSkip\setlength{\cmsTabSkip}{1ex}

\newcommand{\ttbb}{\ensuremath{\cPqt\cPaqt\cPqb\cPaqb}\xspace}
\newcommand{\ttcc}{\ensuremath{\cPqt\cPaqt\cPqc\cPaqc}\xspace}
\newcommand{\ttbj}{\ensuremath{\cPqt\cPaqt\cPqb\cmsSymbolFace{j}}\xspace}
\newcommand{\ttLF}{\ensuremath{\cPqt\cPaqt\cmsSymbolFace{LF}}\xspace}
\newcommand{\ttccLF}{\ensuremath{\cPqt\cPaqt\cPqc\cPaqc\cmsSymbolFace{LF}}\xspace}
\newcommand{\ttjj}{\ensuremath{\cPqt\cPaqt\cmsSymbolFace{jj}}\xspace}
\newcommand{\bb}{\ensuremath{\cPqb\cPaqb}\xspace}

\newcommand{\ee}  {\ensuremath{\Pepm\Pemp}\xspace}
\newcommand{\mumu}{\ensuremath{\PGmpm\PGmmp}\xspace}
\newcommand{\mue} {\ensuremath{\Pepm\PGmmp}\xspace}

\newcommand{\ttW}{\ensuremath{\cPqt\cPaqt\PW}\xspace}
\newcommand{\ttZ}{\ensuremath{\cPqt\cPaqt\PZ}\xspace}
\newcommand{\ttH}{\ensuremath{\cPqt\cPaqt\PH}\xspace}
\newcommand{\VV}{\ensuremath{\PV\PV}\xspace}
\newcommand{\VVV}{\ensuremath{\PV\PV\PV}\xspace}
\newcommand{\ttV}{\ensuremath{\cPqt\cPaqt\PV}\xspace} 

\newcommand{\tW}{\ensuremath{\cPqt\PW}\xspace}

\newcommand{\rbbjj}{\ensuremath{R_{\ttbb/\ttjj}}\xspace}

\newcommand{\xsecttbb}{\ensuremath{\sigma_{\ttbb}}\xspace}
\newcommand{\xsecttjj}{\ensuremath{\sigma_{\ttjj}}\xspace}

\newcommand{\Eff}{\ensuremath{\mathcal{E}}\xspace}
\newcommand{\Acc}{\ensuremath{\mathcal{A}}\xspace}
\newcommand{\FPS}{\ensuremath{\cmsSymbolFace{FPS}}\xspace}
\newcommand{\VPS}{\ensuremath{\cmsSymbolFace{VPS}}\xspace}

\newcommand{\mcNLO}{MG\_a\MCATNLO}

\newcommand{\LumiVal}{\ensuremath{35.9\fbinv}\xspace}

\cmsNoteHeader{TOP-18-002} 
\title{Measurement of the cross section for $\ttbar$ production with additional jets and \cPqb jets in $\Pp\Pp$ collisions at $\sqrt{s}=13\TeV$}
\author{ \LARGE The CMS Collaboration}
\date{\today}

\abstract{Measurements of the cross section for the production of top quark pairs in association with a pair of jets from bottom quarks (\xsecttbb) and in association with a pair of jets from quarks of any flavor or gluons (\xsecttjj) and their ratio are presented. The data were collected in proton-proton collisions at a center-of-mass energy of 13\TeV by the CMS experiment at the LHC in 2016 and correspond to an integrated luminosity of \LumiVal. The measurements are performed in a fiducial phase space and extrapolated to the full phase space, separately for the dilepton and lepton+jets channels, where lepton corresponds to either an electron or a muon. The results of the measurements in the fiducial phase space for the dilepton and lepton+jets channels, respectively, are $\xsecttjj = 2.36\pm 0.02\stat\pm 0.20\syst\unit{pb}$ and $31.0\pm 0.2\stat\pm 2.9\syst\unit{pb}$, and for the cross section ratio $0.017\pm 0.001\stat\pm 0.001\syst$ and $0.020\pm 0.001\stat\pm 0.001\syst$. The values of \xsecttbb are determined from the product of the \xsecttjj and the cross section ratio, obtaining, respectively, $0.040\pm 0.002\stat\pm 0.005\syst\unit{pb}$ and $0.62\pm 0.03\stat\pm 0.07\syst\unit{pb}$. These measurements are the most precise to date and are consistent, within the uncertainties, with the standard model expectations obtained using a matrix element calculation at next-to-leading order in quantum chromodynamics matched to a parton shower.}

\hypersetup{pdfauthor={CMS Collaboration},pdftitle={Measurement of the cross section for top quark-antiquark pairs production with additional jets and b jets in pp collisions at sqrt(s) = 13 TeV},pdfsubject={CMS},pdfkeywords={CMS, top quark, cross section, b quark}}

\maketitle 

\section{Introduction}
\label{sec:intro}  
The production of top quark-antiquark pairs ($\ttbar$) in association with two inclusive jets, \ttjj, where j denotes jets produced from the fragmentation of quarks of any flavor (\cPqu, \cPqd, \cPqc, \cPqs, \cPqb) or from gluons, and the special case of $\ttbar$ production in association with a \bb pair (\ttbb) are interesting from theoretical and experimental points of view.
Even though calculations at next-to-leading order (NLO) in quantum chromodynamics (QCD) are available for both processes~\cite{Bevilacqua:2014qfa,PhysRevD.84.114017, Jezo:2018yaf}, they suffer from large uncertainties in the choice of factorization and renormalization scales~\cite{Bredenstein:2008zb}, because of the presence of two very different scales, the top quark mass ($m_{\cPqt}$) and the jet transverse momentum (\pt), that both play a role in these processes.
The experimental measurements of the proton-proton ($\Pp\Pp$) production cross sections for $\Pp\Pp \to \ttjj~(\xsecttjj)$ and $\Pp\Pp \to \ttbb~(\xsecttbb)$ provide a useful test of NLO QCD calculations.

The \ttbb process is also a dominant background for different measurements of the Higgs boson and the top quark properties~\cite{Sirunyan:2018egh,Sirunyan2018,Sirunyan2017,2018345,PhysRevD.98.052005}.
The Yukawa coupling of the Higgs boson to the top quark is one of the relevant properties; its measurement probes the consistency of the standard model (SM) Higgs sector.
The top quark Yukawa coupling can be probed directly in the associated production of the Higgs boson with $\ttbar$ (\ttH). 
This process was recently observed by the ATLAS and CMS Collaborations ~\cite{ttHCMS,ttHATLAS}. 
Both experiments also reported searches for \ttH production in the \bb decay channel of the Higgs boson~\cite{Sirunyan:2018mvw,PhysRevD.97.072016}.
A challenge in this decay channel is the irreducible nonresonant background from $\ttbb$ production.

In this paper, we present measurements of \xsecttbb, \xsecttjj, and their ratio ($\rbbjj$) in $\Pp\Pp$ collisions at a center-of-mass energy of 13\TeV, using data corresponding to an integrated luminosity of \LumiVal collected in 2016 with the CMS detector at the LHC. 
The \xsecttjj cross section and $\rbbjj$ ratio measurements are separately performed in the dilepton (\ee, \mumu and \mue) and lepton+jets ($\Pepm$+jets and $\PGmpm$+jets) channels, by means of binned maximum likelihood fits to the measured \cPqb tagging discriminant distribution of additional jets in events with $\ttbar$ candidates. 
The \xsecttbb is determined by multiplying the obtained \xsecttjj by the $\rbbjj$.
The cross section ratio has a smaller relative systematic uncertainty than the absolute cross section measurements, due to the partial cancellation of uncertainties. 

As described in Ref.~\cite{Bevilacqua:2014qfa}, the kinematic range of the theoretical predictions for $\ttbar$ production in association with jets plays a fundamental role in the reliability of the fixed-order estimation of the $\rbbjj$ ratio.
At jet $\pt$ values higher than $\approx$40\GeV, the stability of the perturbative expansion of $\xsecttjj$, and $\xsecttbb$ can be lost and resummation of higher-order effects must be considered.
For this reason, the analysis is performed in two phase space regions (defined in Section~\ref{sec:signal}) with different jet $\pt$ requirements.
These two measurements can improve the modeling of the $\ttbb$ process as well as the understanding of the related theoretical uncertainties.

Previous measurements of the $\ttbb$ to $\ttjj$ cross section ratio at center-of-mass energies of 8 and 13\TeV have been performed by ATLAS~\cite{atlas:ttbb8TeV,Aaboud:2018eki} and CMS~\cite{cms:ttbb8TeV, cms:TOP-12-041, 078433dfbc4e4b04a600834663f932ea}. 
The 8\TeV results by ATLAS and CMS are compatible with the SM prediction within the experimental uncertainties of 33 and 26\%, respectively.
The ATLAS Collaboration measures at 13\TeV the production of $\ttbar$ pairs as a function of the \cPqb jet multiplicity in the dilepton (lepton+jets) channel with an uncertainty of 13 (17)\%.
These measurements are compatible with theoretical predictions within the uncertainties.
The CMS measurement at 13\TeV in the dilepton channel reports a value for this ratio that is 1.8 times larger than the NLO SM prediction. However, the total uncertainty in this measurement is 32\% and the statistical significance of this deviation is therefore only around two standard deviations. 
The latest measurement of the inclusive $\ttbb$ cross section performed by CMS~\cite{CMS-PAS-TOP-18-011} in the fully hadronic final state reports a cross section slightly higher but consistent with the prediction with a total uncertainty of 29\%.

This paper presents the most precise measurements of the $\ttbb$ to $\ttjj$ cross section ratio and the inclusive $\ttjj$ cross section to date by means of an improved fit method and a data set about 15 times larger than the previous CMS measurement~\cite{ 078433dfbc4e4b04a600834663f932ea} performed in the dilepton channel.
Additionally, this is the first measurement of the cross sections for  $\ttbar$ events with additional jets and \cPqb jets in the lepton+jets channel performed using the CMS data at 13\TeV.
This decay channel has a much higher branching fraction than the dilepton channel, increasing the available number of events to be analyzed.

The paper is organized as follows. 
A brief description of the CMS detector is given in Section~\ref{sec:CMSDetector}.
Details of the event simulation of the $\ttbar$ and other SM processes, together with the theoretical calculation of the $\ttbar$ cross section, are given in Section~\ref{sec:datamc}.
Section~\ref{sec:signal} contains the definitions of the fiducial and full phase space regions and the $\ttjj$ event categorization.   
The event selection and the methods used to measure the $\ttjj$ cross sections and the $\ttbb$ to $\ttjj$ cross section ratios are discussed in Sections~\ref{sec:sel} and \ref{sec:xsec}, respectively.
A complete description of the systematic uncertainties is given in Section~\ref{sec:syst}.
Finally, the cross section measurements are reported in Section~\ref{sec:results} and a summary is provided in Section~\ref{sec:summ}.  

\section{The CMS detector}
\label{sec:CMSDetector}
The central feature of the CMS apparatus is a superconducting solenoid of 6\unit{m} internal diameter, providing a magnetic field of 3.8\unit{T}. 
Within the solenoid volume are a silicon pixel and strip tracker, a lead tungstate crystal electromagnetic calorimeter (ECAL), and a brass and scintillator hadron calorimeter, each composed of a barrel and two endcap sections. 
Forward calorimeters extend the pseudorapidity ($\eta$) coverage provided by the barrel and endcap detectors. 
The electron momentum is estimated by combining the energy measurement in the ECAL with the momentum measurement in the tracker. 
The momentum resolution for electrons is generally better in the barrel region than in the endcaps, and depends on the bremsstrahlung energy emitted by the electron as it traverses the material in front of the ECAL~\cite{Khachatryan:2015hwa}. 
Muons are detected in gas-ionization chambers embedded in the steel flux-return yoke outside the solenoid. Muons are measured in the range $\abs{\eta} < 2.4$, with detection planes made using three technologies: drift tubes, cathode strip chambers, and resistive plate chambers.
The efficiency to reconstruct and identify muons is greater than 96\%.
Matching muons to tracks measured in the silicon tracker results in a relative transverse momentum resolution, for muons with \pt up to 100\GeV, of 1\% in the barrel and 3\% in the endcaps. The \pt resolution in the barrel is better than 7\% for muons with \pt up to 1\TeV~\cite{Sirunyan:2018fpa}.

The global event reconstruction (also called particle-flow (PF) event reconstruction~\cite{Sirunyan:2017ulk}) aims to reconstruct and identify each individual particle in an event, with an optimized combination of all subdetector information. In this process, the identification of the particle type (photon, electron, muon, charged or neutral hadron) plays an important role in the determination of the particle direction and energy.
For each event, hadronic jets are clustered from the reconstructed particles using the infrared and collinear-safe anti-\kt algorithm~\cite{Cacciari:2008gp, Cacciari:2011ma} with a distance parameter of 0.4. 
Jet momentum is determined as the vectorial sum of all particle momenta in the jet, and is found from simulation to be within 5 to 10\% of the true momentum over the whole \pt spectrum and detector acceptance.
The missing transverse momentum vector \ptvecmiss is computed as the negative vector sum of the transverse momenta of all the PF candidates in an event, and its magnitude is denoted as \ptmiss~\cite{Sirunyan:2019kia}. The \ptvecmiss is modified to account for corrections to the energy scale of the reconstructed jets in the event.
The candidate vertex with the largest value of summed physics-object $\pt^2$ is taken to be the primary $\Pp\Pp$ interaction vertex.
The physics objects are the jets and the associated $\ptmiss$.

Events of interest are selected using a two-tiered trigger system~\cite{Khachatryan:2016bia}.
The first level, composed of custom hardware processors, uses information from the calorimeters and muon detectors to select events at a rate of around 100\unit{kHz} within a time interval of less than 4\mus.
The second level, known as the high-level trigger, consists of a farm of processors running a version of the full event reconstruction software optimized for fast processing, and reduces the event rate to around 1\unit{kHz} before data storage.

A more detailed description of the CMS detector, together with a definition of the coordinate system used and the relevant kinematic variables, can be found in Ref.~\cite{Sirunyan:2017ulk}.

\section{Signal and background simulation} 
\label{sec:datamc} 
The signal and background processes are simulated using Monte Carlo (MC) techniques. 
The \ttbar signal sample is generated by \POWHEG (v2)~\cite{Nason:2004rx,Frixione:2007vw,Alioli:2010xd,Campbell:2014kua} at NLO and combined with the parton shower (PS) and underlying event (UE) simulation from \PYTHIA{}8 (v8.219)~\cite{Sjostrand:2014zea} using the CUETP8M2T4~\cite{CMS-PAS-TOP-16-021} tune. A second \POWHEG sample employs \HERWIG{}++~\cite{Bahr2008} (v2.7.1) using the tune EE5C~\cite{Seymour:2013qka}.
In these samples additional (\cPqb) jets beyond real emission at NLO are generated by the parton shower.
The proton structure is described by the parton distribution function (PDF) set NNPDF3.0~\cite{Dulat:2015mca}.
To compare with an alternative theoretical prediction, \MGvATNLO (v2.2.2) (\mcNLO) is used to generate \ttbar events in the five-flavor scheme (5FS)~\cite{Frixione:2008yi, Re:2010bp}.
The \textsc{madspin}~\cite{madspin} package is used to incorporate the correct treatment of the decay particles preserving spin correlation effects. 
In addition to the \ttbar pair, up to two additional partons are simulated at NLO and matched using the FXFX algorithm~\cite{Frederix2012}, and the PS simulation is performed by \PYTHIA{}8, denoted as \mcNLO+ \PYTHIA{}8 5FS [FXFX]. 
Detailed explanations of the different parameters in the $\ttbar$ simulations can be found in~\cite{CMS-PAS-TOP-16-021}.
The $\ttbar$ simulations are normalized to their inclusive cross section, $832~^{+20}_{-29}\,(\text{scale})\pm{35}$~(PDF and strong coupling $\alpS)\unit{pb}$, calculated with the \textsc{top++} v2.0 program~\cite{topplusplus} to next-to-next-to-leading order (NNLO) in perturbative QCD, including soft-gluon resummation at next-to-next-to-leading-logarithmic order~\cite{Cacciari:2011hy, Baernreuther:2012ws, Czakon:2012zr, Czakon:2012pz, Beneke:2011mq, topNNLO}.

The production of single top quarks in the \tW channel is simulated at NLO with the \POWHEG generator in the 5FS, and normalized to the cross sections calculated at NNLO~\cite{Kidonakis:2012rm}.
The ``diagram removal'' (DR) scheme~\cite{Frixione:2008yi} is used to account for the interference with ttbar production.
The $t$-channel single top quark production is simulated at NLO in the four-flavor scheme (4FS). 
The $s$-channel single top quark production is generated at NLO in the 4FS with \mcNLO, which is also used for the PS simulation.
The \mcNLO generator with the MLM merging scheme~\cite{Alwall:2007fs} is used for the simulation at LO of \PW{}+jets and \PZ{}+jets production, and the samples are normalized to the inclusive cross sections calculated at NNLO~\cite{Li:2012wna}.
The background contribution from $\ttbar$ production in association with a Higgs boson ($\ttH$) is generated with \POWHEG, while $\ttbar$ production in association with a \PW and a \PZ bosons (referred to as \ttV) is generated at NLO using \mcNLO. These samples are normalized to the cross sections at NLO~\cite{Alwall_2014,Campbell:2011bn}.
Both generators used for the $\ttbar$ production in association with bosons are interfaced with \PYTHIA{}8.  
All MC simulations with top quark production assume a top quark mass of 172.5\GeV~\cite{ATLAS:2014wva}.

Diboson production ($\PW\PW$, $\PW\PZ$, and $\PZ\PZ$, referred to as \VV) is simulated at leading order using \PYTHIA{}8, and normalized to the cross sections calculated at NNLO for the $\PW\PW$ sample~\cite{PhysRevLett.113.212001} and NLO for the $\PW\PZ$ and $\PZ\PZ$ samples~\cite{Campbell:2010ff}. 
Triboson production (referred to as \VVV) is generated at NLO using \mcNLO, and \PYTHIA{}8 is used for simulation. 
The \PYTHIA{}8 generator is also used to generate the QCD multijet background events.
The CMS detector response is simulated using \GEANTfour (v.9.4)~\cite{geant}.
The simulations include multiple $\Pp\Pp$ interactions per bunch crossing (pileup).
The simulated events are weighted, depending on the number of pileup interactions, to reproduce the observed pileup distribution in data (on average, 23 collisions per bunch crossing).

\section{Definitions of the \texorpdfstring{$\ttjj$}{ttbar jet-jet} categories and regions of phase space} 
\label{sec:signal}
Events with a $\ttbar$ pair and at least two additional jets in simulation are categorized further using the flavor of the particle-level jets, found from the MC generator information.
The particle-level jets are obtained by clustering final-state particles with a mean lifetime greater than 30\unit{ps} (except neutrinos) using the anti-\kt algorithm with a distance parameter of 0.4. 
The flavor of the particle-level jet is identified by the ghost-matching technique~\cite{Cacciari:2008gn}. 
The ghost-matched clusters are formed with final-state particles supplemented by hadrons containing bottom and charm quarks (called bottom and charm hadrons) that do not have further bottom or charm hadrons as daughter particles.
The momentum of each bottom and charm hadron is artificially reduced to zero in order to avoid affecting the observable particle-level jet momentum. 
If a particle-level jet contains bottom or charm hadrons, it is assigned to the corresponding flavor and called a {\cPqb} or {\cPqc} jet.
Otherwise, it is considered as a particle-level light-flavor jet (from a gluon, or {\cPqu}, {\cPqd}, or {\cPqs} quark). 
Particle-level jets are identified as products of the $\ttbar$ decay if the bottom hadron used for the flavor assignment belongs to the simulation history of any of the top quarks.

The different $\ttjj$ categories are based on the flavor of the particle-level jets that accompany the $\ttbar$ system: $\ttbb$ with at least two additional particle-level \cPqb jets; $\ttbj$ with one additional particle-level \cPqb jet and at least one additional particle-level \cPqc or light-flavor jet; $\ttcc$ with at least two additional particle-level \cPqc jets; and $\ttLF$ with at least two additional particle-level light-flavor jets or one particle-level light-flavor jet and one \cPqc jet.
The $\ttbj$ final state mainly originates from the merging of two particle-level \cPqb jets or from one of the particle-level \cPqb jets failing the following acceptance requirements.
The particle-level jets have to satisfy $\abs{\eta} < 2.5$ and $\pt >$ 30 (20)\GeV in the dilepton (lepton+jets) channel.
The different $\pt$ cut for the dilepton and lepton+jet channels allows the definition of different phase space regions, providing additional information for testing the theoretical predictions.
The category $\ttjj$ comprises all categories defined above, \ie all events with at least two additional particle-level jets, regardless of their flavor.
The $\ttbar$ events that do not belong to any of the $\ttjj$ categories, \eg, events with only one additional particle-level jet, are treated as background.
The particle-level leptons originating from the decays of the top quarks, are defined using the anti-\kt clustering algorithm with a distance parameter of 0.1 to account for final-state radiated photons, and are required to fulfill  $\abs{\eta} < 2.4$ and \pt $ >$ 20 (30)\GeV for the dilepton (lepton+jets) channel. 

Measurements of the inclusive cross sections, $\xsecttbb$, $\xsecttjj$, and their ratio are reported both in the fiducial (``visible'') phase space (VPS) and the full phase space (FPS). 
Simulated events are defined as signal events in the VPS if they contain particle-level leptons and particle-level jets satisfying the above acceptance requirements, and fulfill the following separate criteria. 
In the dilepton channel, $\ttbb$ ($\ttjj$) events are considered in the VPS if they have two particle-level leptons and at least four particle-level jets, including at least four (two) particle-level \cPqb jets. In a similar way, for the lepton+jets channel, $\ttbb$ ($\ttjj$) events are considered in the VPS if they have only one particle-level lepton and at least six particle-level jets, including at least four (two) particle-level \cPqb jets. 
The FPS is defined as the \ttbb (\ttjj) final state with all \ttbar decay channels and at least two additional particle-level \cPqb (any) jets that do not originate from top quark decays. 
The selection criteria are summarized in Table~\ref{tab:ps}. 
The measurements performed in the FPS facilitate comparisons to QCD calculations at NLO.

\begin{table}[tbh]
\topcaption{
Summary of the requirements for a simulated event to be in the fiducial (VPS) and full (FPS) phase space regions for the $\ttbb$ and $\ttjj$ categories in the dilepton and lepton+jets channels.
Details of the particle-level definitions are described in the text.
The symbol $\ell$ denotes a lepton (\Pe or \PGm).
}
\label{tab:ps}
\centering{
\cmsTable
{
\begin{tabular}{lcccc}
Channel    &  Jet $\pt$ & Phase space    &   $\ttbb$                                                 &  $\ttjj$                       \\  \hline
\multirow{2}{*}{Dilepton} & \multirow{2}{*}{$>$30\GeV} & VPS                     &  $\ell\ell$ + $\ge$ 4 jets (4 \cPqb jets)                     &    $\ell\ell$  + $\ge$ 4 jets (2 \cPqb jets)   \\
& & FPS                        &  $\ttbar$ + $\ge$ 2 \cPqb jets (not from $\ttbar$) & $\ttbar$ + $\ge$ 2 jets (not from $\ttbar$)    \\
\vspace{-2mm}& & & & \\
\multirow{2}{*}{Lepton+jets} & \multirow{2}{*}{$>$20\GeV} & VPS                     &  $\ell$ + $\ge$ 6 jets (4 \cPqb jets)                       & $\ell$ + $\ge$ 6 jets (2 \cPqb jets)     \\
& & FPS                        &  $\ttbar$ + $\ge$ 2 \cPqb jets (not from $\ttbar$) & $\ttbar$ + $\ge$ 2 jets (not from $\ttbar$)    \\
\end{tabular}
}
}
\end{table}

\section{Event selection}
\label{sec:sel}
The measurements are performed independently in the dilepton and lepton+jets final states of the ${\ttbar}$ decay.
A combination of single-lepton and dilepton triggers with specific transverse momentum requirements is applied to filter events in the dilepton channel.
In the $\ee$ and $\mumu$ channels, the charged lepton with the highest \pt is required to pass the trigger with $\pt > 23$ (17)\GeV for the electron (muon), and the charged lepton with the lowest \pt must have $\pt > 12$ (8)\GeV. 
Events in the $\mue$ channel are selected requiring a trigger with either one electron with $\pt > 12\GeV$ and one muon with $\pt > 23\GeV$, or one electron with $\pt > 23\GeV$ and one muon with $\pt > 8\GeV$. 
Additionally, events selected by a set of single-lepton triggers with one electron (muon) with $\pt > 27$ (20)\GeV are assigned to one of the three dilepton combinations. 
Events for the lepton+jets channel are required to pass a single-electron (-muon) trigger with $\pt > 32$ (24)\GeV. 
The higher-$\pt$ thresholds for the single-lepton triggers used in the lepton+jets channel reduce the contribution of processes with leptons from bottom/charm hadron decays, electrons from misidentified jets, etc.
Any overlap between the dilepton and lepton+jets channels is avoided by requiring a different number of leptons in each decay channel.

Leptons originating from the \PW boson in top quark decays are expected to be isolated. 
Therefore, isolation criteria are applied to the reconstructed and selected electrons~\cite{Khachatryan:2015hwa} and muons~\cite{Sirunyan:2018fpa}. 
A relative isolation variable, the \pt sum of charged and neutral hadrons, and photons in a cone of $\Delta R = \sqrt{\smash[b]{(\Delta \phi)^2 +(\Delta \eta)^2}}$ around the direction of the lepton divided by the lepton \pt, is used, where $\Delta \phi$ and $\Delta \eta$ are the azimuthal angle and pseudorapidity differences, respectively, between the directions of the lepton and the other particle. 
The relative isolation parameter defined in a cone of $\Delta R < 0.3$ (0.4) is required to be lower than 0.06 (0.15) for electrons (muons).

The energy of the jets reconstructed using the anti-\kt jet clustering algorithm~\cite{Cacciari:2011ma,Cacciari:2008gp} is corrected for the pileup contributions using the average energy density deposited by neutral particles in the event. 
The charged hadron subtraction mitigates event by event the effect of tracks coming from pileup on the transverse energy of the jet.
Jet energy corrections are also applied as functions of the jet \pt and $\eta$~\cite{Khachatryan:2016kdb}. 
The \cPqb jets are identified using the combined secondary vertex (CSVv2) \cPqb tagging algorithm~\cite{Sirunyan:2017ezt}. 
Two different operating points for the CSVv2 algorithm are chosen, based on the expected background composition of each decay channel. 
The operating point selected for the dilepton (lepton+jets) channel corresponds to an efficiency for correctly identifying \cPqb jets of about 70 (50)\% and a probability of about 1 (0.1)\% to misidentify a light-flavor jet as a \cPqb jet~\cite{Sirunyan:2017ezt}. 
A tighter operating point in the lepton+jets channel is chosen to increase the rejection of background processes with multijet final states. 
Since the MC simulation does not reproduce exactly the \cPqb tagging discriminant distribution observed in data, scale factors are applied to the simulated events in order to correct its shape. 
The scale factors, derived using a tag-and-probe technique in enriched regions for both \cPqb and light-flavor jets, are applied as a function of the jet \cPqb tagging discriminant, $\pt$, and $\eta$~\cite{Sirunyan:2017ezt}.

In the dilepton channel, events are selected to have exactly two oppositely charged leptons with $\abs{\eta} < 2.4$ and $\pt > 25\GeV$ for the higher-\pt (leading) lepton and $\pt > 20\GeV$ for the subleading one. 
Additional requirements include: the invariant mass of the dilepton system $m_{\ell^{+}\ell^{-}}\, > 20\GeV$, four or more jets with $\pt > 30\GeV$ and $\abs{\eta} < 2.4$, and at least two jets to be identified as a \cPqb jet. 
Further selection criteria are applied for the same-flavor lepton channels ($\ee$ and $\mumu$) to reject events from \PZ{}+jets as follows: $\ptmiss > 40\GeV$ and $\abs{m_{\ell^{+}\ell^{-}}\,- m_{\PZ}} > 15\GeV$, where $m_{\PZ}$  is the $\PZ$ boson mass of 91\GeV~\cite{PDG2018}.
With these requirements, events with two jets from the $\ttbar$ process and at least two additional jets are selected.

Events in the lepton+jets channel are required to have exactly one isolated electron (or muon) with \pt $ >$ 35 (30)\GeV and $\abs{\eta} < 2.1$ (2.4), thus avoiding events selected for the dilepton channel, and at least six reconstructed jets with $\pt > 30\GeV$ and $\abs{\eta} < 2.4$, where at least two of them are identified as \cPqb jets. 
This requirement reflects the presence of four jets from the $\ttbar$ process, plus at least two additional jets. 
A higher reconstructed-jet $\pt$ requirement in comparison with the $\pt$ criterion for particle-level jets is applied so that all events passing the reconstruction selection criteria will be in the VPS.
Besides the differences in the trigger requirements for the dilepton and lepton+jets channels, a higher-$\pt$ requirement for the selected leptons reduces the contribution from processes such as the QCD multijet background.  

Table~\ref{tab:yields} lists the expected final numbers of data events in the dilepton and lepton+jets channels. 
The numbers are given for the different $\ttbar$(+jets) categories, assuming $\sigma_{\ttbar}=832 \unit{pb}$ from an NNLO calculation and the individual sources of background (from MC simulation), normalized to an integrated luminosity of $\LumiVal$.
The category ``$\ttbar$ others'' corresponds to the background contribution from $\ttbar$ events that pass the event selection but do not belong to any of the $\ttjj$ categories.
The background contributions from \PW{}+jets and QCD multijets in the dilepton channel and \VVV in the lepton+jets channel are negligible in the corresponding final state. 
The expected sample composition contains more than 75 (84)\% of $\ttjj$ signal events in the dilepton (lepton+jets) channel. 
The small background contributions from non-$\ttbar$ processes are estimated from simulation.
The number of observed events for each channel is also given in Table~\ref{tab:yields}, and is consistent with the expected number within the uncertainties.

\begin{table}[hbt]
\topcaption{Expected and observed numbers of events in the dilepton and lepton+jets channels after applying the event selection.  
The results are given for the different $\ttbar$(+jets) categories, the individual sources of background (from MC simulation), normalized to an integrated luminosity of $\LumiVal$, and the observed number from data.
The uncertainties quoted for each MC contribution include all the systematic uncertainties described in Section~\ref{sec:syst}.
}
\label{tab:yields}
\centering{    
\begin{tabular}{lr@{~$\pm$~}lr@{~$\pm$~}l}   
Source (MC simulation) & \multicolumn{2}{c}{Dilepton} & \multicolumn{2}{c}{Lepton+jets} \\\hline
$\ttbb$                     & 327    & 23   & 2470   & 180   \\
$\ttbj$                     & 1103   & 62   & 3820   & 170   \\
$\ttcc$                     & 353    & 23   & 1627   & 74    \\
$\ttLF$                     & 10\,860 & 580  & 26\,900 & 1300 \\[\cmsTabSkip]
Total $\ttjj$               & 12\,640 & 590  & 34\,800 & 1400 \\[\cmsTabSkip]
\ttbar\ others              & 3740   & 180  & 4180   & 190   \\
Single top quark            & 500    & 100  & 1460   & 160   \\
\PW{}+jets & \multicolumn{2}{c}{\NA}             & 250    & 35    \\
\PZ{}+jets                      & 50     & 18   & 78     & 23    \\
\VV                          & 3      & 2    & 10     & 5     \\
\VVV                         & 1      & 1    & \multicolumn{2}{c}{\NA}   \\
QCD multijet &    \multicolumn{2}{c}{\NA}    & 220    & 130   \\
$\ttH$                      & 54     & 22   & 230    & 130   \\
\ttV                   & 86     & 16   & 381    & 55    \\[\cmsTabSkip]
Total signal and background & 17\,100 & 620  & 41\,600 & 1400  \\[\cmsTabSkip]
Data & \multicolumn{2}{c}{16\,167} & \multicolumn{2}{c}{39\,819}   \\
\end{tabular}
}
\end{table}

\section{Cross section measurement} 
\label{sec:xsec}
The selected events described in Section~\ref{sec:sel} contain at least four (six) jets in the dilepton (lepton+jets) channel where at least two of them must be identified as \cPqb jets. 
In order to identify the origin of the selected jets (either from a top quark decay or not), two different approaches are followed, based on the kinematic properties of each final state.
In the dilepton channel, the two \cPqb jets with the largest values of the \cPqb tagging discriminant provided by the CSVv2 algorithm originate from a top quark decay in 85 (23)\% of selected $\ttjj$ ($\ttbb$) events~\cite{078433dfbc4e4b04a600834663f932ea}, as determined from simulation. 
Therefore, the jets with the third- and fourth-largest \cPqb tagging discriminant values are considered as additional jets.
The identification of the origin of the jets in the lepton+jets channel is more complex because the relatively large number of jets (at least six) leads to ambiguities in the jet assignment. 
Therefore, a jet assignment based only on the \cPqb tagging discriminant is insufficient and a kinematic fit~\cite{KinFitD0} is applied.

The event inputs for the kinematic fit algorithm are the four-momenta of the selected lepton and jets, whether a jet is identified as a \cPqb jet, and $\ptmiss$.
The kinematic fit algorithm constrains the momentum of the aforementioned objects to the hypothesis that two top quarks of the same mass are produced, each one decaying to a bottom quark and a \PW boson. 
The \PW boson decay products are also constrained to an invariant mass of $80.4\GeV$~\cite{PDG2018}. 
The last constraint in the kinematic fit requires that the jets associated with \cPqb quarks from top quark decays must be identified as \cPqb jets as well.
The algorithm assigns a $\chi^2$ value~\cite{PhysRevD.97.112003} to each solution according to the goodness of the fit of each jet permutation.
The solution selected is the one with the lowest $\chi^2$ value. 
This guarantees that the selected jet permutation is the most compatible with the $\ttbar$ process.
All jets that are not included in the selected solution (i.e., not coming from the top quark or \PW boson decay) are considered as additional jets for the lepton+jets channel.
This assumption leads to a correct identification of at least one additional jet in 70 (50)\% of the cases for the $\ttjj$ ($\ttbb$) category.
An efficiency of 40 (12)\% is reached for the correct identification of both additional (\cPqb) jets.

\begin{figure}[!htb]
\centering{
\includegraphics[width=0.48\textwidth]{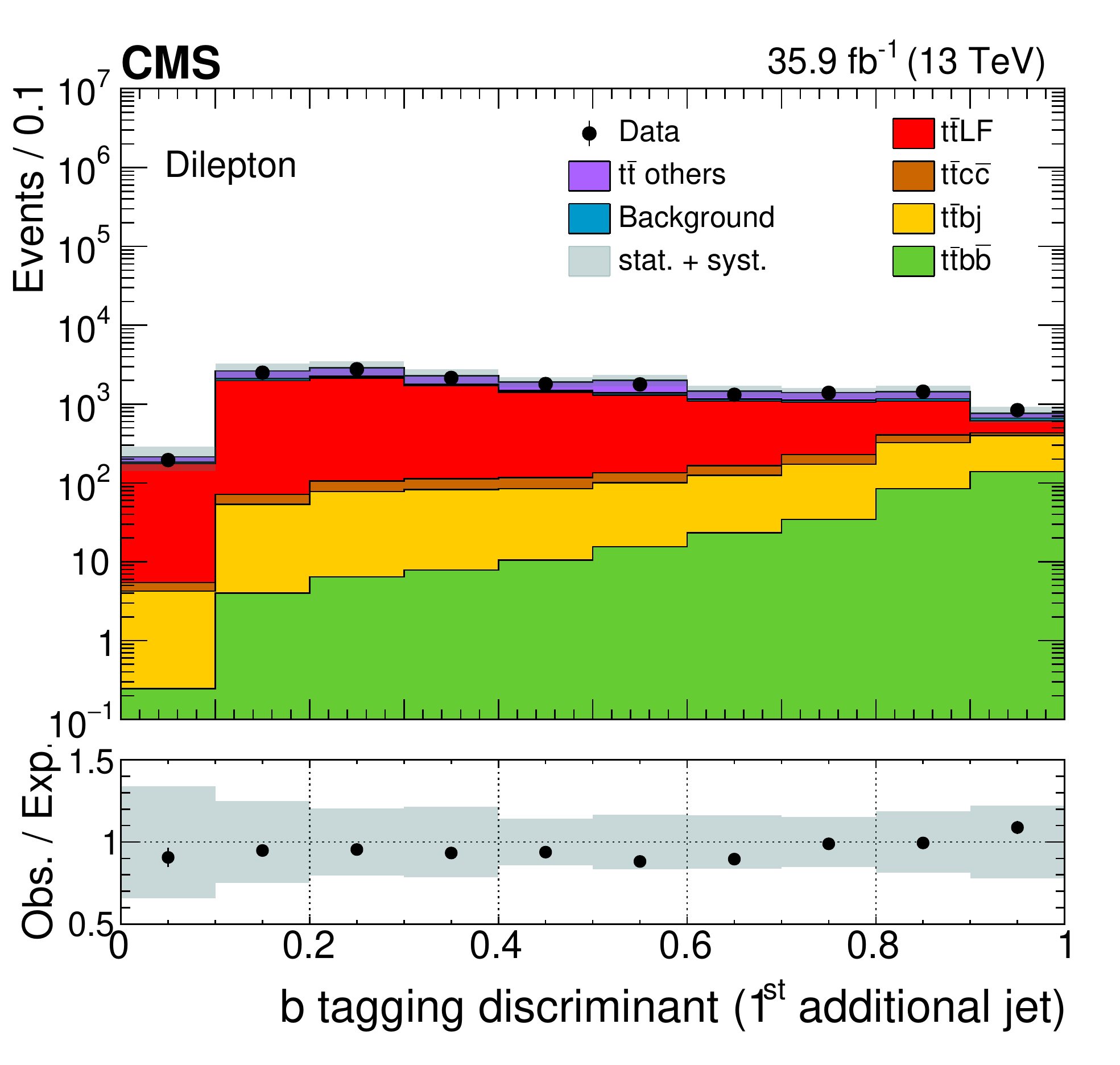} \includegraphics[width=0.48\textwidth]{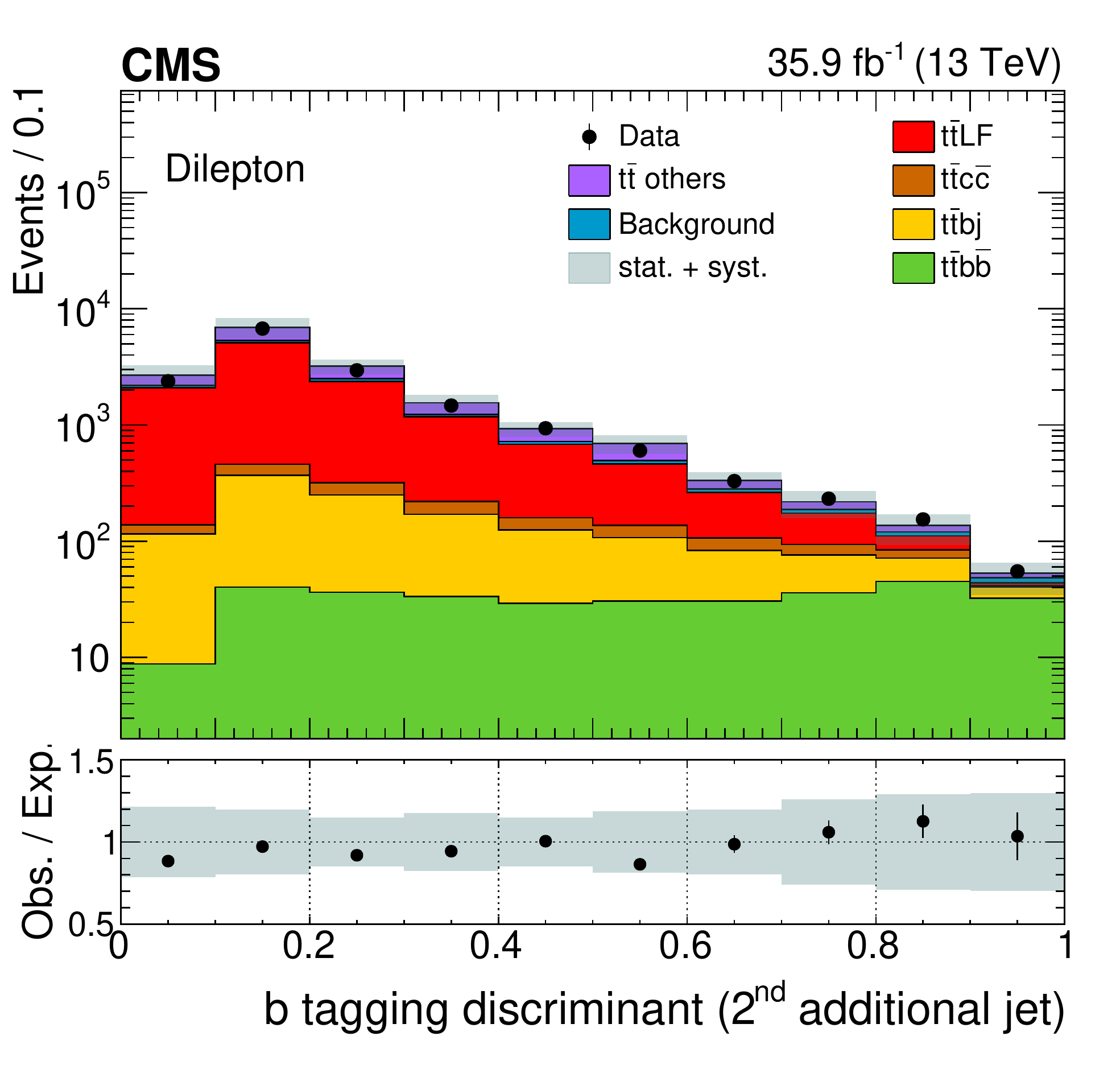}
\includegraphics[width=0.48\textwidth]{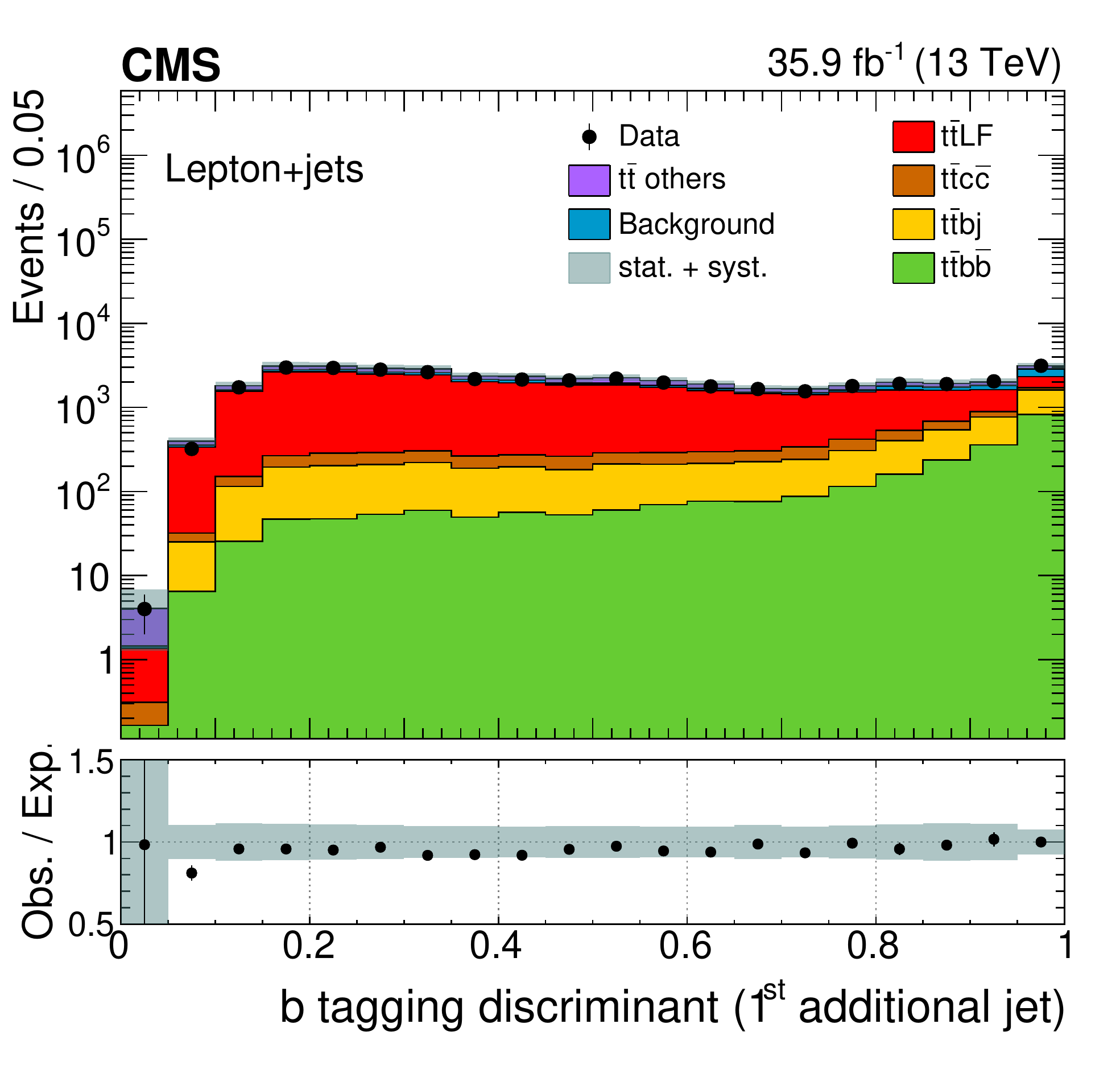} \includegraphics[width=0.48\textwidth]{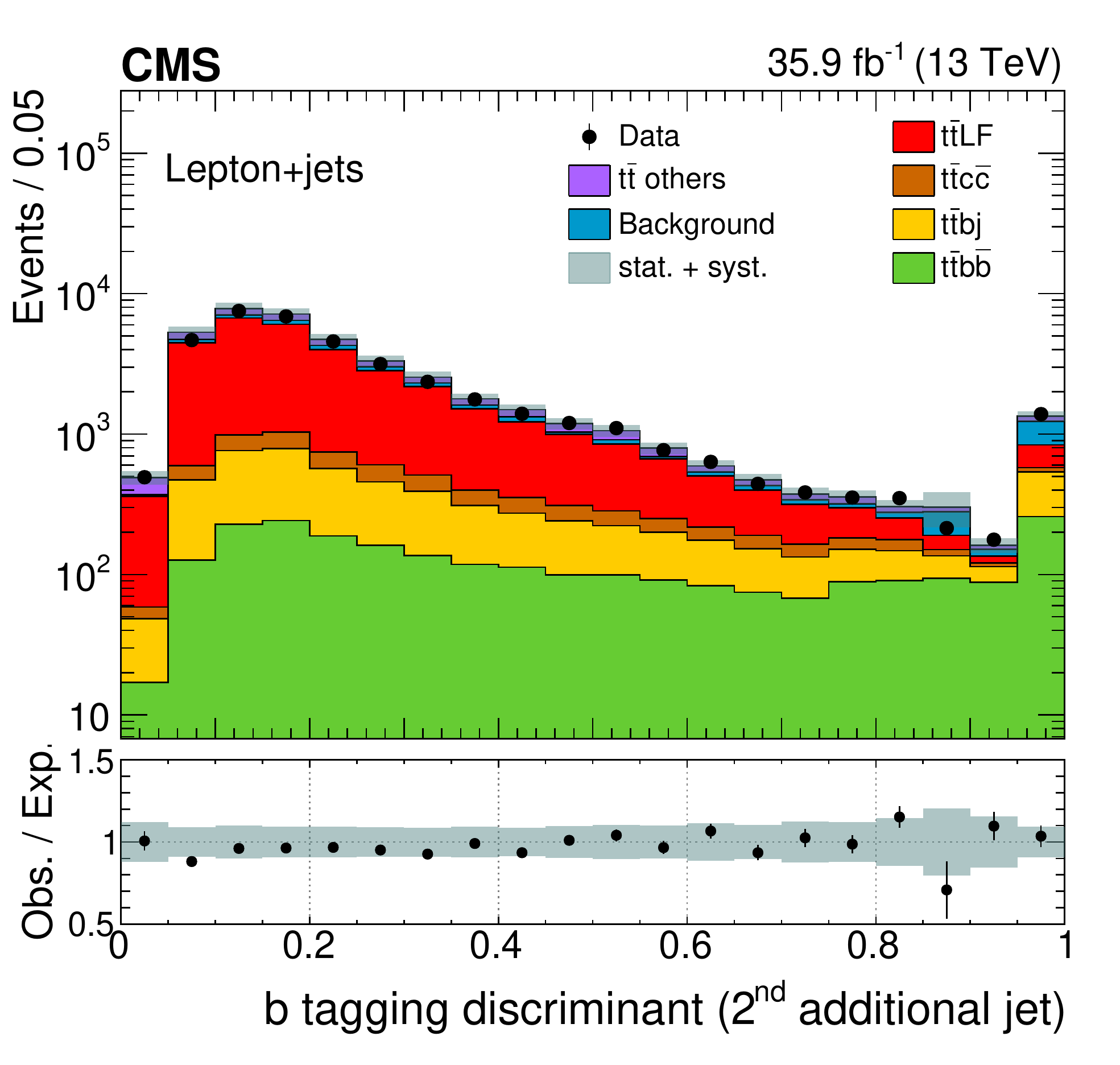}
}
\caption{The \cPqb tagging discriminant distribution from data (points) for the first (left) and second (right) additional jet for the dilepton (upper) and lepton+jets (lower) channels in decreasing order of the \cPqb tagging discriminant value after event selection, and the predicted distributions for the signal and background from simulation (shaded histograms). The contributions of single top quark, \ttV, $\ttH$, and the non-top quark processes are merged in the background (blue) entry.
The $\ttbb$ process is located in the bottom part of the stack due to its low contribution in comparison with the other entries.
The lower panels display the ratio of the data to the expectations. The grey bands display the combination of the statistical and systematic uncertainties.
}
\label{fig:addjetcsv}
\end{figure}

After identifying the origin of the jets of the $\ttbar$ system, the additional jets in the event are arranged in decreasing order of the \cPqb tagging discriminant value.
Only the first two additional jets (those with the highest \cPqb tagging discriminant value) are kept for further analysis.
Figure~\ref{fig:addjetcsv} shows the \cPqb tagging discriminant distribution from data for the first (left) and second (right) additional jets in the dilepton (upper) and lepton+jets (lower) channels, along with the predictions from the MC simulations.
The values of the \cPqb tagging discriminant for the first two additional jets allows one to distinguish between the different $\ttjj$ categories.

\begin{figure}[hbtp]
\centering{
\includegraphics[width=0.4\textwidth]{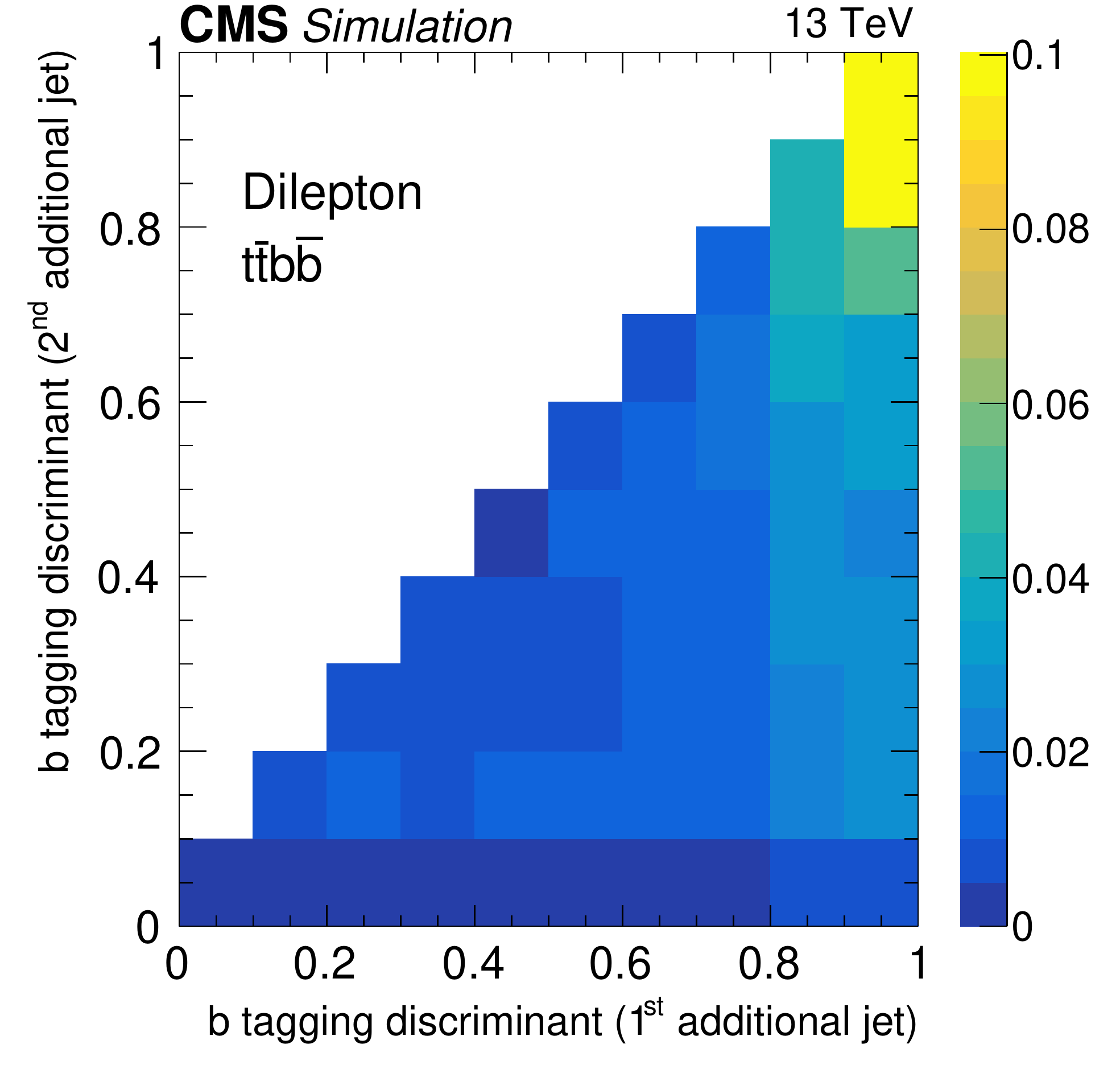}\hspace{1cm}\includegraphics[width=0.4\textwidth]{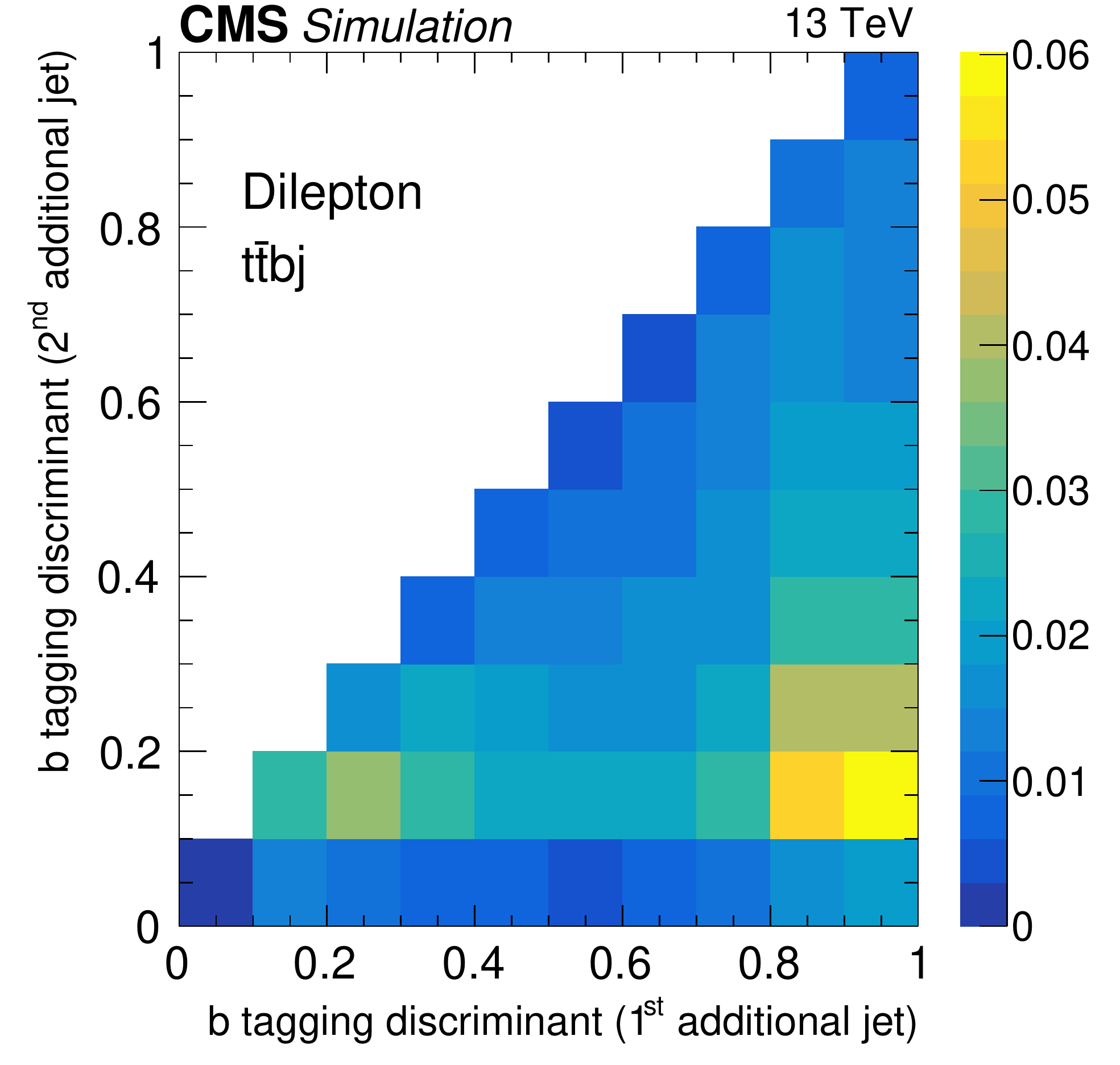}
\includegraphics[width=0.4\textwidth]{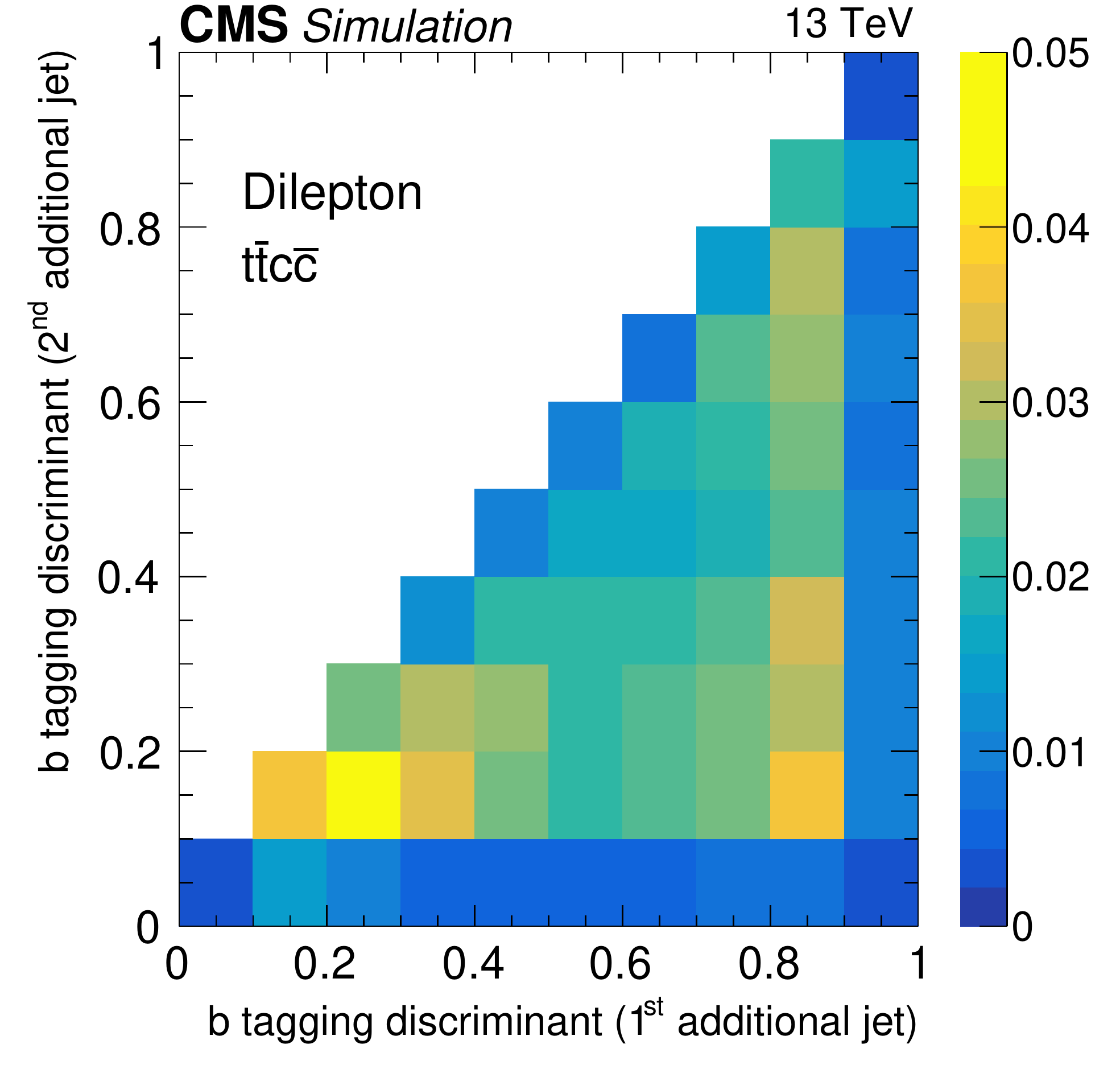}\hspace{1cm}\includegraphics[width=0.4\textwidth]{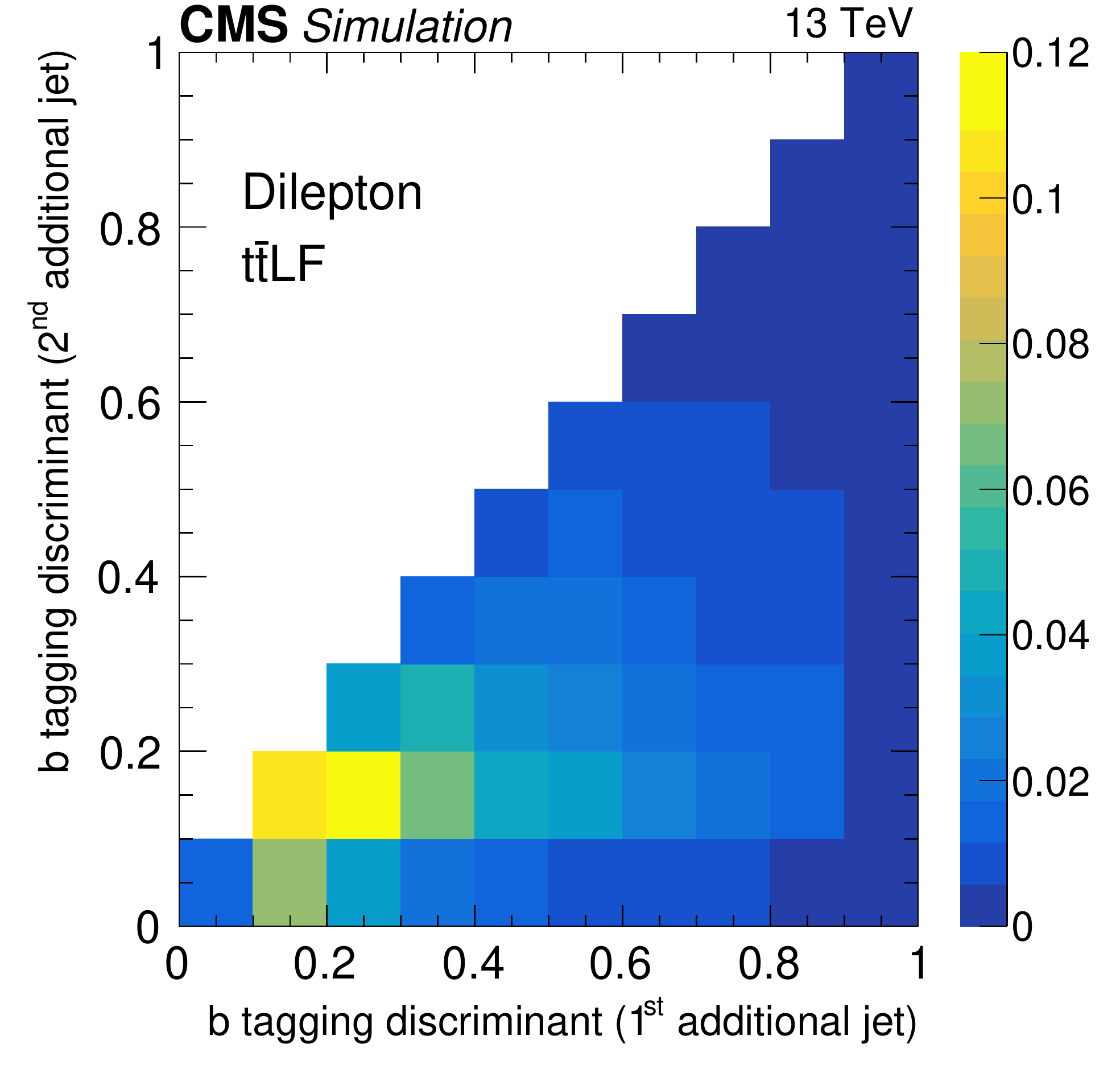}
}
\caption{Two-dimensional distributions of the \cPqb tagging discriminant for the first and second additional jets in the dilepton channel shown separately for different flavors of the additional jets:
$\ttbb$ (upper left),
$\ttbj$ (upper right),
$\ttcc$ (lower left) and $\ttLF$ (lower right).
The number of entries is normalized to unity.
The histograms are obtained from the \POWHEG MC simulation.}
\label{fig:btagshape2D}
\end{figure}

\begin{figure}[hbt]
\centering{
\includegraphics[width=0.4\textwidth]{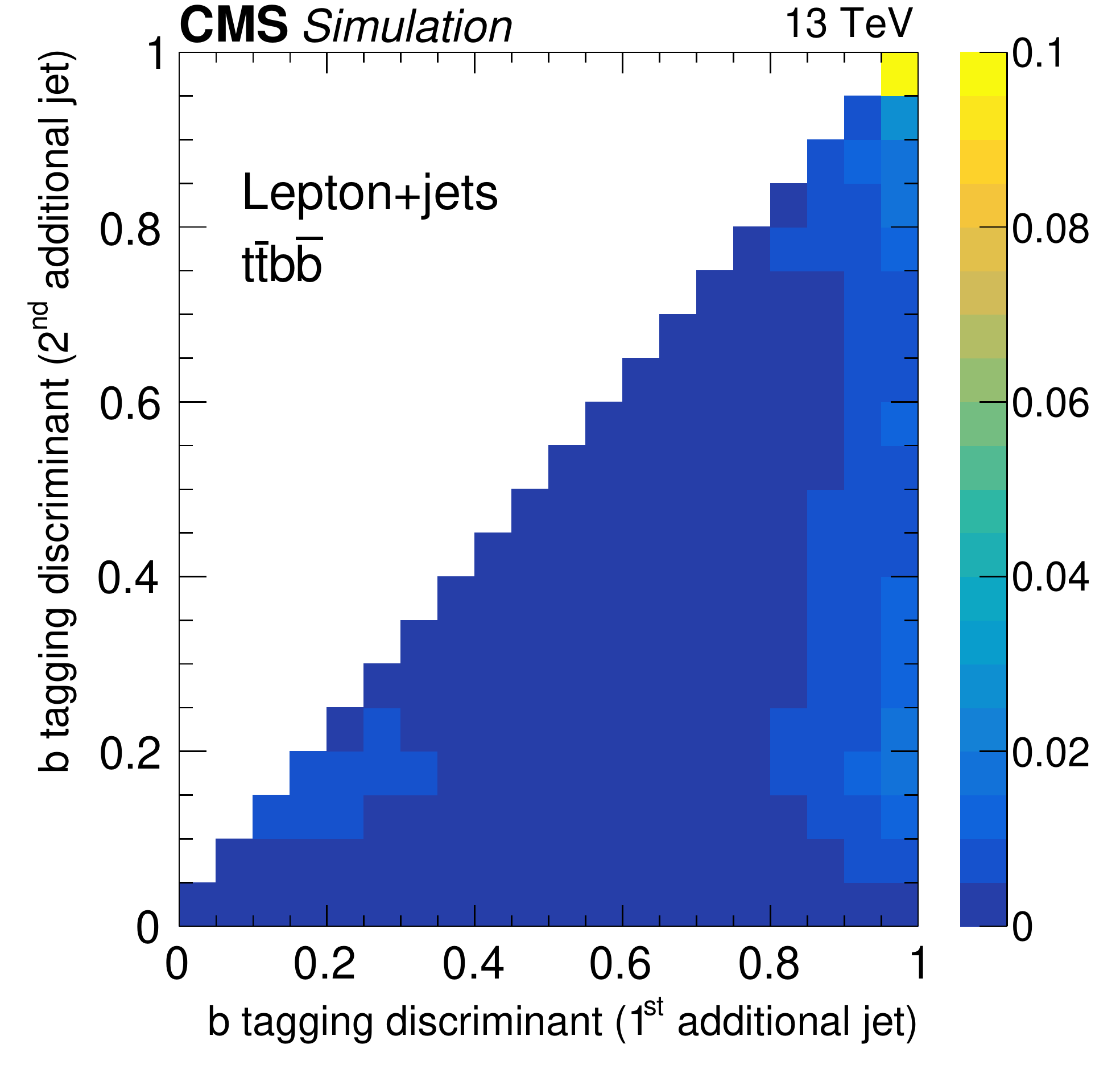}\hspace{1cm}\includegraphics[width=0.4\textwidth]{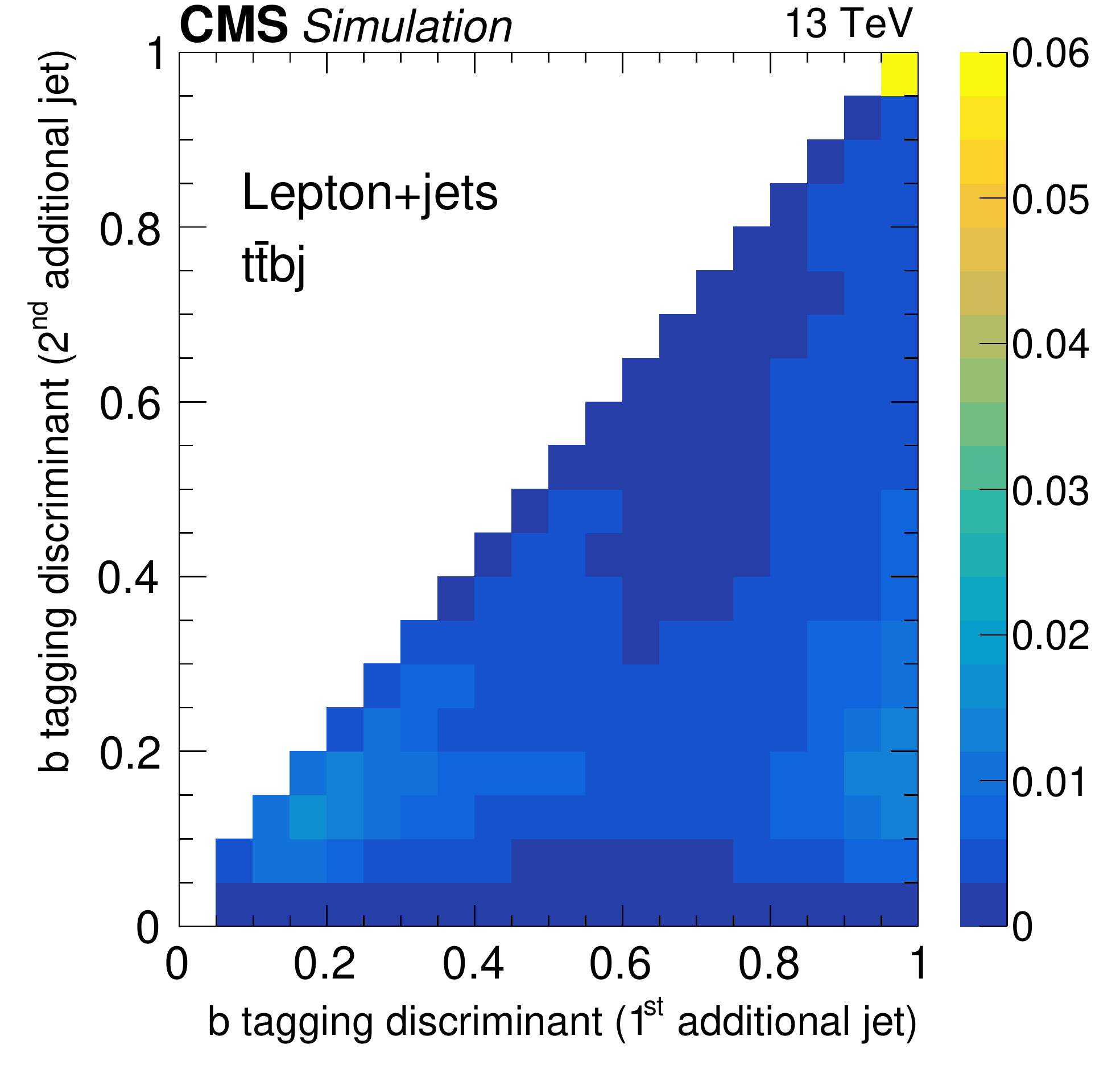}
\includegraphics[width=0.4\textwidth]{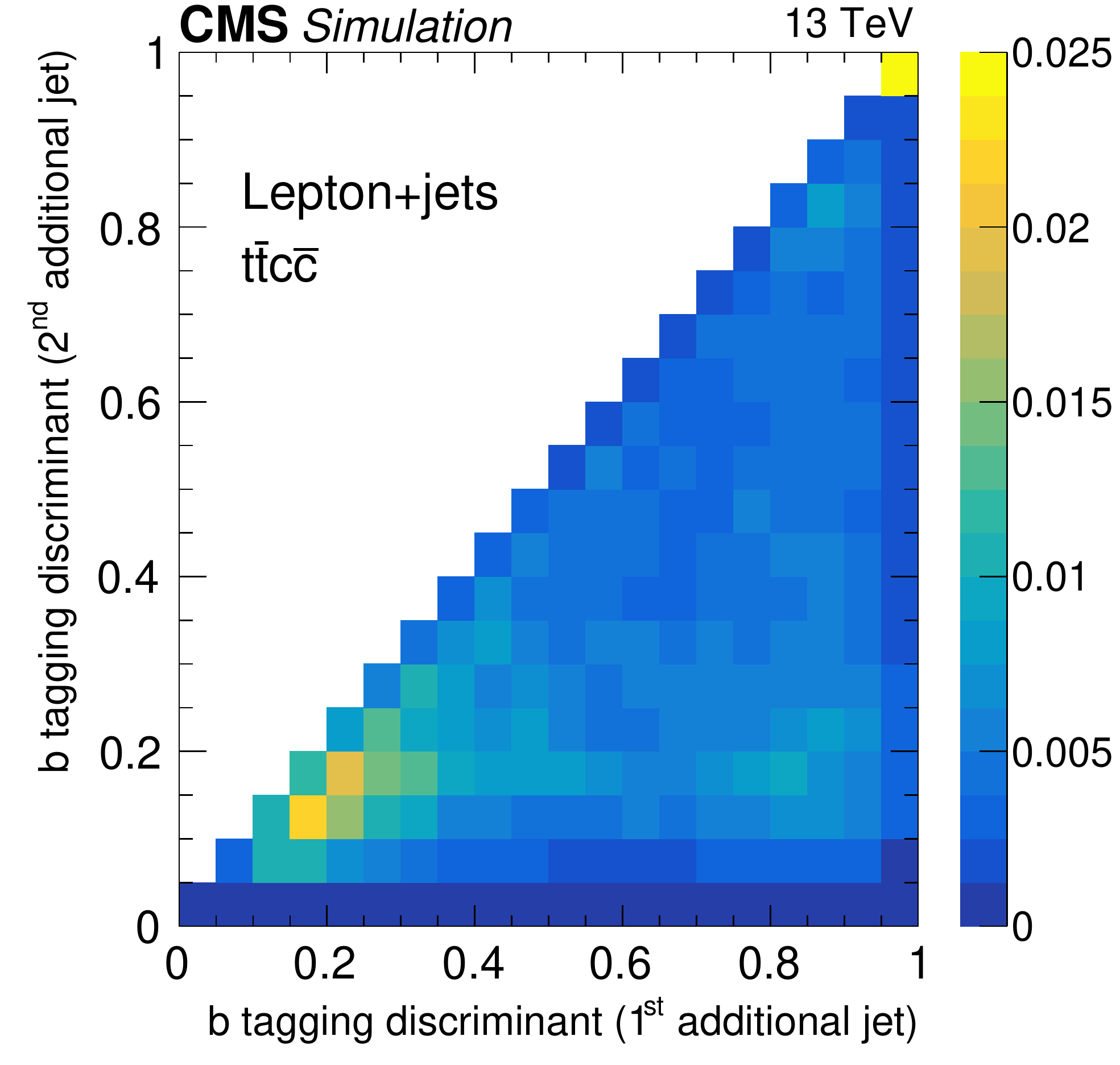}\hspace{1cm}\includegraphics[width=0.4\textwidth]{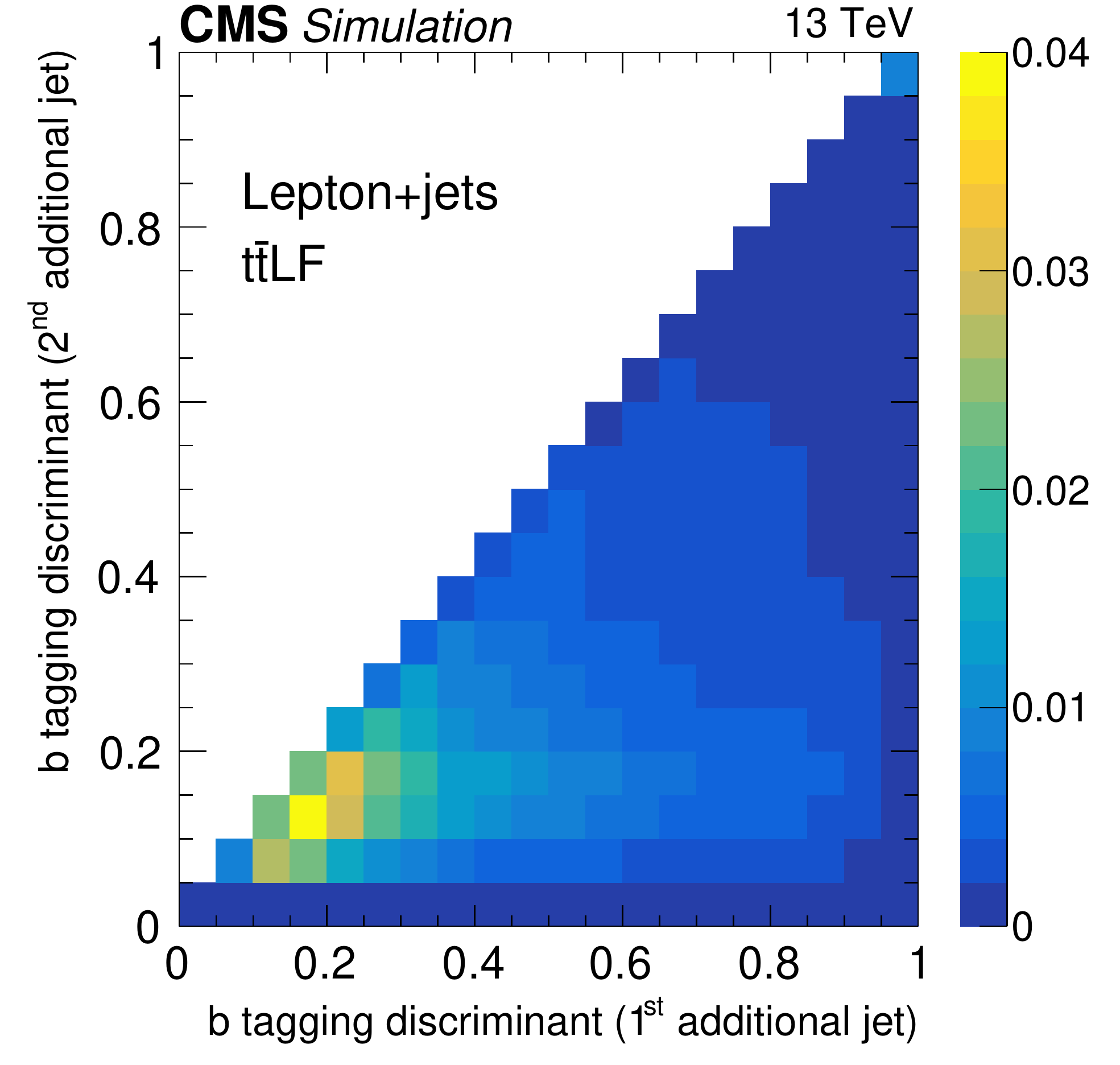}
}
\caption{Two-dimensional distributions of the \cPqb tagging discriminant for the first and second additional jets in the lepton+jets channel shown separately for different flavors of the additional jets:
$\ttbb$ (upper left),
$\ttbj$ (upper right),
$\ttcc$ (lower left) and $\ttLF$ (lower right).
The number of entries is normalized to unity.
The histograms are obtained from the \POWHEG MC simulation.}
\label{fig:csvnorm}
\end{figure}

Exploiting the separation power of the \cPqb tagging discriminant, the $\ttbb$ to $\ttjj$ cross section ratio and the absolute $\ttjj$ cross section are simultaneously extracted using a binned maximum likelihood fit.
Figure~\ref{fig:btagshape2D} (\ref{fig:csvnorm}) shows the two-dimensional (2D) distributions of the \cPqb tagging discriminant from simulation separately for the $\ttbb$, $\ttbj$, $\ttcc$, and $\ttLF$ events in the dilepton (lepton+jets) channel.
The dilepton and lepton+jets decay channels are fitted as two independent measurements because of the different phase space definitions and different background compositions. 
Then, within the dilepton and the lepton+jets channels, the different final states are fitted simultaneously.

The following procedure is applied to each final state for both the dilepton and the lepton+jets channels.
A 2D joint distribution is formed from the \cPqb tagging discriminant values of both additional jets. 
The distribution is binned into $10{\times}10$ ($20{\times}20$) equidistant intervals for the dilepton (lepton+jets) channel.

The likelihood depends on the two parameters of interest (POI) in the VPS, the $\ttjj$ cross section ($\xsecttjj^\VPS$) and the $\ttbb$ to $\ttjj$ cross section ratio ($\rbbjj^\VPS$), and on nuisance parameters affecting the kinematic distributions to be considered as part of the systematic uncertainties described in Section~\ref{sec:syst}.
It is constructed as the product over all bins of the 2D joint distribution of a Poisson probability with a mean defined in each bin by:

\begin{equation}
\label{eq:one}
\mathcal{M}{\scriptstyle \left(\xsecttjj^{\VPS},~\rbbjj^{\VPS}\right)} = \left(\xsecttjj^{\VPS} \lumi \Eff_{\ttjj}\right)  \mathrm{pdf}_{\ttjj}^\mathrm{\;norm} +  \left( \frac{\xsecttjj^{\VPS}}{\xsecttjj^\mathrm{MC}}\right)  \mathrm{pdf_{\ttbar\,others}} +  \mathrm{pdf_{bkg}},
\end{equation}

where $\mathrm{pdf_{bkg}}$ is the probability density function (pdf) for each bin of the full set of backgrounds presented in Table~\ref{tab:yields}, including $\ttW$, $\ttZ$, and $\ttH$.
Those contributions are fixed in the fit to their estimated yields from the MC simulation.
Potential deviations of $\xsecttjj^\VPS$ with respect to the expectation from simulation ($\xsecttjj^\mathrm{MC}$) are propagated to the ``$\ttbar$ others'' component. 
Therefore, $\mathrm{pdf_{\ttbar\,others}}$ enters the likelihood function corrected by the $\xsecttjj^{\VPS}/\xsecttjj^\mathrm{MC}$ term, where the constant parameter $\xsecttjj^{\mathrm{MC}}=163$ (290)\unit{pb} for the dilepton (lepton+jets) channel is taken from MC simulation. 
The remaining term in Eq.~(\ref{eq:one}) contains the total integrated luminosity, $\lumi$, the efficiency for selecting $\ttjj$ events, $\Eff_{\ttjj}$, and the pdf normalized to unity for each bin of the $\ttjj$ category, $\mathrm{pdf}_{\ttjj}^\mathrm{norm}$. The latter parameter is defined as:
\begin{equation}
\label{eq:two}
\mathrm{pdf}_{\ttjj}^\mathrm{\;norm} =\left[ \rbbjj^{\VPS} \left(\frac{\Eff_{\ttbb}}{\Eff_{\ttjj}} \right) \mathrm{pdf}_{\ttbb}^\mathrm{\;norm} + R_{\ttbj/\ttjj}\,\mathrm{pdf}_{\ttbj}^\mathrm{\;norm} + R_{\ttccLF/\ttjj}\,\mathrm{pdf}_{\ttccLF}^\mathrm{\;norm}\right],
\end{equation}
where $\mathrm{pdf}_{\ttbb}^\mathrm{norm}$, $\mathrm{pdf}_{\ttbj}^\mathrm{norm}$, and $\mathrm{pdf}_{\ttccLF}^\mathrm{norm}$ are the normalized pdfs of the $\ttbb$, $\ttbj$, and combined $\ttcc$ and $\ttLF$ ($\ttccLF$) categories.
The parameter $R_{\ttbj/\ttjj}$~ is the expected ratio of the number of $\ttbj$ events to the number of $\ttjj$ events. 
This ratio is expressed as a function of the POI $\rbbjj^{\VPS}$:
\begin{equation}
\label{eq:three}
R_{\ttbj/\ttjj} = \rbbjj^{\VPS} \left(\frac{\Eff_{\ttbb}}{\Eff_{\ttjj}}\right) R_{\ttbj/\ttbb}^\mathrm{MC}.
\end{equation}
The parameter $R_{\ttbj/\ttbb}^\mathrm{MC}$ is the expected cross section ratio from MC simulation of the $\ttbj$ and $\ttbb$ processes. 
This value is taken to be constant ($R_{\ttbj/\ttbb}^\mathrm{MC}=1.5$), considering the correlation between the definition of those two $\ttbar$ categories. 
The parameter $\Eff_{\ttbb}$ is the efficiency for selecting $\ttbb$ events. 
The event selection efficiencies, defined as the number of events after the full event selection divided by the number of events in the VPS, are 25 (18)\% for $\ttbb$ and 10 (5)\% for $\ttjj$ in the dilepton (lepton+jets) channel. 
The quantity $R_{\ttccLF/\ttjj}$ is the ratio of the number of $\ttccLF$ events to the number of $\ttjj$ events, which satisfies the unitarity requirement of $R_{\ttccLF/\ttjj} + R_{\ttbb/\ttjj} + R_{\ttbj/\ttjj} = 1$.
The binned maximum likelihood fit includes the effect of the systematic uncertainties, described in Section~\ref{sec:syst}, as nuisance parameters. 

The $\ttbb$ to $\ttjj$ cross section ratio and the absolute $\ttjj$ cross section in the FPS are extrapolated from the results in the VPS by: 
\begin{equation}
\label{eq:FromVtoF}
\rbbjj^{\FPS} = \frac{\Acc_{\ttjj}}{\Acc_{\ttbb}} \rbbjj^{\VPS} ;\qquad \xsecttjj^{\FPS} = \frac{\xsecttjj^{\VPS}}{\Acc_{\ttjj}\;\mathcal{B}}.   
\end{equation}
Here, $\Acc_{\ttbb}$ and $\Acc_{\ttjj}$ are the acceptances for the $\ttbb$ and $\ttjj$ processes, respectively, taken from the nominal MC simulation, and $\mathcal{B}$ is the branching fraction of the $\ttbar$ pair to each decay channel considered.
The acceptance values, defined as the ratio between the number of events in the visible and full phase space (including decays into $\tau$ leptons in the full phase space) are 14 (31)\% for $\ttbb$ and 15 (25)\% for $\ttjj$ in the dilepton (lepton+jets) channel.
The total branching fraction for $\ttbar$ decays in the dilepton (lepton+jets) channel is 0.104 (0.436)~\cite{PDG2018}, including decays into the three lepton flavors.

Finally, the absolute $\ttbb$ cross section $\xsecttbb$ is calculated by multiplying the POI $\rbbjj$ and $\xsecttjj$ in each phase space region. 
The total uncertainty in $\xsecttbb$ is estimated by considering the uncertainties in each POI and their correlation after the fit.

\section{Systematic uncertainties}
\label{sec:syst} 
The systematic uncertainties are incorporated as nuisance parameters in the maximum likelihood fit used to extract both POI: the cross section ratio of $\ttbb$ to $\ttjj$ production and the absolute $\ttjj$ production cross section. 
The relevance of extracting the cross section ratio is that several systematic uncertainties cancel, specifically those related to the common normalization in both processes, such as the integrated luminosity, trigger efficiencies, lepton identification, and energy scale, as well as the jet energy scale (JES) and resolution (JER). 

There are two main categories of uncertainties: those from the detector performance and signal efficiency, and those from the modeling of the signal.
All the uncertainties coming from the detector performance are included in the maximum likelihood fit as shape-changing nuisance parameters. 
The theoretical uncertainties from modeling are treated as rate-changing nuisance parameters. 
This consideration is based on studies that show shape-changing nuisance parameters in distributions with large statistical fluctuations (as the $\ttbar$ MC samples for modeling uncertainties) produce incorrect constraints in the post-fit uncertainties.
The studies also show that, under these conditions, the rate-changing nuisance parameters cover any expected shape variation.

The trigger uncertainties are about 1\% for all the lepton flavor combinations in both channels. 
The systematic uncertainty in the lepton identification efficiency is calculated by varying the correction factor for the efficiency within its uncertainty, as derived from \PZ boson candidates as a function of the lepton $\eta$ and \pt, and also taking into account the different phase space between \PZ boson and \ttbar events. 
Lepton energy scale variations are also considered.

The effect of JES and JER variations is estimated as follows. 
The uncertainty produced by the JER is obtained by smearing the jet energy in simulation by a value dependent on $\pt^\text{jet}$ and $\eta_\text{jet}$.
In the case of the JES, the energies of all jets in the event are scaled up and down according to the 25 independent variations described in Ref.~\cite{Khachatryan:2016kdb}.
The effect of the energy scale on the \ptmiss is considered by propagating the variations of jet momenta and the jet energy smearing to its value.
The JES uncertainties are also propagated to the \cPqb tagging discriminant efficiencies.

The systematic uncertainties associated with the \cPqb tagging efficiency for \cPqb jets and light-flavor jets are studied separately by varying their values within their uncertainties~\cite{Sirunyan:2017ezt}.
The variations are divided into different categories that include two statistical effects and three jet flavor effects.
The statistical uncertainties account for fluctuations in data and simulation, assuming a linear dependence for the effect on the slope and a quadratic dependence for the overall shift of the \cPqb tagging discriminant distribution.
The uncertainty in the contamination by light- and heavy-flavor jets, respectively, in the control regions used to determine the heavy- and light-flavor jet efficiencies is estimated by varying the contamination fractions by 20\%.
The \cPqc jet scale factors are assumed to be unity with an uncertainty twice as large as the \cPqb jet tagging scale factors.

The uncertainties in the amount of initial- and final-state radiation, ISR and FSR, respectively, are considered by varying the corresponding scale parameters by factors of 2 and $\sqrt{2}$~\cite{Skands:2014pea}. 
The matrix element (ME) to PS matching uncertainty is evaluated by varying the model parameter by $h_\text{damp} = (1.58~^{+0.66}_{-0.59}$)~$m_{\cPqt}$, where $m_{\cPqt}$ is the top quark mass, fixed at $172.5\GeV$~\cite{CMS-PAS-TOP-16-021}.
This parameter is used in the \POWHEG simulation to control the matching of the jets from the ME calculations to those from the \PYTHIA{}8 PS within its uncertainty.

The effect of the tune used by \PYTHIA{}8 to simulate the underlying event is evaluated by varying the tune parameters according to their uncertainties \cite{Skands:2014pea}.

The $\ttbb$ cross section strongly depends on the choice of the factorization ($\mu_\mathrm{F}$) and renormalization ($\mu_\mathrm{R}$) scales in the ME calculations, which are estimated by making use of a weighting scheme implemented in \POWHEG to vary the scales by a factor of two up and down with respect to their reference values $\mu_\mathrm{F} = \mu_\mathrm{R} = \sqrt{\smash[b]{m^2_{\cPqt}+\pt^2}}$, and $\pt$ is the transverse momentum of the top quark. 
The $\mu_\mathrm{F}$ and $\mu_\mathrm{R}$ scales are assumed uncorrelated.

Previous measurements of the differential $\ttbar$ cross section by CMS~\cite{PhysRevD.97.112003} have shown a mismodeling of the \pt distribution of the top quark.
The current \POWHEG{} + \PYTHIA{}8 simulation has a harder top quark \pt distribution than the one obtained from data. 
The uncertainty (labeled top-\pt) is estimated by reweighting the \pt distribution in the simulation to match the data.

The uncertainty associated with the $\ttbj/\ttbb$ ratio (parameter $R_{\ttbj/\ttbb}^\mathrm{MC}$ in Eq.(\ref{eq:two})) is estimated by varying the central value obtained from the MC prediction by $\pm10\%$.
This variation covers the differences obtained by comparing the predictions from the different $\ttbar$ MC simulations.

The uncertainties affecting the acceptances for $\ttbb$ and $\ttjj$ events, used in the extrapolation to the $\FPS$, are not constrained in order to avoid any prior in the theoretical model.
The theoretical uncertainties described above are evaluated, in addition to the effects of the PDF and the color reconnection (CR).
The PDF uncertainty is estimated using the 100 individual uncertainties and $\alpS$ in the NNPDF3.0 set~\cite{Ball:2014uwa}, following the prescription of PDF4LHC~\cite{Butterworth:2015oua}. 
The nominal CR model uses a scheme based on multiple parton interactions with early resonance decays (ERD) turned off.
The uncertainty from CR is estimated by employing three alternative schemes: with the ERD switched on, a QCD-inspired procedure~\cite{Christiansen:2014847}, and a gluon-move scheme~\cite{Argyropoulos:2014zoa}.
The uncertainty in the $\ttbar$ branching fraction is not included in the extrapolation to the FPS since its effect is negligible.

The number of pileup interactions in data is estimated from the measured bunch-by-bunch luminosity multiplied by the total inelastic $\Pp\Pp$ cross section.  
The uncertainty assigned to the pileup simulation is obtained by varying the inelastic cross section of $69.2\unit{mb}$ by $\pm4.6\%$~\cite{LUM-17-001}.

To estimate the background uncertainties, several rate-changing nuisance parameters have been included, even if they do not affect the measurement significantly.
Different uncertainties depending on the experimental precision reached in the measurement of the background cross sections are applied. 
For the dilepton channel, the background uncertainties are assessed conservatively by varying each contribution (\PZ{}+jets, \VV, \VVV, \ttV, single top quark) by 30\% to cover the uncertainty in the production cross sections, and detector performance uncertainties, of these background processes.
In the lepton+jets channel, a 15\% systematic uncertainty in the single top quark background normalization is included, in addition to the other detector performance uncertainties, and 10\% for the \PW{}+jets background normalization. 
For the QCD multijet background, a variation of 50\% beyond the statistical uncertainty is assumed to cover the systematic variations in this process. 
Finally, for the smaller backgrounds \VV, \ttV, and \ttH, uncertainties of 10, 20, and 60\% are used, respectively.

Since the POI are sensitive to the finite number of available simulated events in the \ttbb process (referred to as ``simulated sample size'' in Table~\ref{tab:syst}), the effective number of unweighted events in each histogram bin is added as a nuisance parameter with a Poisson constraint on the likelihood function.
Finally, the uncertainty in the integrated luminosity for the 2016 data-taking period is considered, corresponding to 2.5\%~\cite{LUM-17-001}.

All the systematic uncertainties are considered as fully correlated between the different final states of each channel, except the trigger uncertainties, which are treated as uncorrelated.

To show the effect of each nuisance parameter, we evaluate the impact of their contributions to the uncertainty in the POI. 
The impact of a nuisance parameter on a fit parameter is defined as the shift of the fit parameter from its post-fit value while fixing the nuisance parameter to $\pm$1 standard deviation from the post-fit value, with all other parameters profiled as normal. 
The systematic uncertainties in $\rbbjj$ and $\xsecttjj$ for the VPS are shown in Table~\ref{tab:syst}.
The total systematic uncertainty in the cross sections ratio ($\ttjj$ cross section) for the VPS is 8.0 (8.8)\% for the dilepton channel and 5.5 (10)\% for the lepton+jets channel.
The JER, some variations of the JES, and the \cPqb tagging uncertainties for the \cPqc-flavor jets are some of the most constrained uncertainties in the fit.

\begin{table}[!htb]
\topcaption{Summary of the individual contributions to the systematic uncertainty in the $\rbbjj$ and $\xsecttjj$ measurements for the VPS. 
The uncertainties are given as relative uncertainties. 
Some sources include a linear (lin.) or quadratic (quad.) dependency on the fluctuations in data and simulation.  
The statistical uncertainty in the result is given for comparison.
}
\centering{
\begin{tabular}{lcccc}
\multirow{2}{*}{Source} & \multicolumn{2}{c}{$\rbbjj^{\VPS}$ [\%]} & \multicolumn{2}{c}{$\xsecttjj^{\VPS}$ [\%]}\\[\cmsTabSkip]
& Dilepton & Lepton+jets & Dilepton & Lepton+jets \\ \hline
\multicolumn{5}{c}{Lepton uncertainties}\\[\cmsTabSkip]
Trigger                   & $<$0.1  & 0.2    &  1.0  & 0.5  \\
Lepton identification     & 0.6  & 0.2    &  1.1  & 1.3  \\
Lepton energy scale       & \NA   & $<$0.1 & \NA    & 0.1  \\[\cmsTabSkip]
\multicolumn{5}{c}{Jet uncertainties}\\[\cmsTabSkip]
Jet energy resolution (JER) & 0.4 & 0.3 & 0.3    & 0.7  \\
Jet energy scale (JES)      & 1.5 & 1.2 & 2.9    & 3.6  \\[\cmsTabSkip]
\multicolumn{5}{c}{b tagging uncertainties}\\[\cmsTabSkip]
\cPqc-flavor \cPqb tag (lin.)     &  2.2 & 2.0 & 1.0    & 0.3  \\
\cPqc-flavor \cPqb tag (quad.)     &  0.7 & 1.2 & 0.3    & 0.2  \\
Heavy-flavor \cPqb tag          &  4.0 & 0.1 & 0.5    & 0.9  \\
Heavy-flavor \cPqb tag (lin.)  &  0.9 & 0.4 & 1.5    & 0.5  \\
Heavy-flavor \cPqb tag (quad.) &  2.0 & 0.3 & 1.5    & 0.8  \\
Light-flavor \cPqb tag          &  4.9 & 0.9 & 5.5    & 4.9  \\
Light-flavor \cPqb tag (lin.)  &  0.1 & 0.2 & 0.3    & 1.1  \\
Light-flavor \cPqb tag (quad.) &  0.7 & 0.7 & 0.1    & 1.4  \\[\cmsTabSkip]
\multicolumn{5}{c}{Theoretical uncertainties}\\[\cmsTabSkip]
Initial-state radiation (ISR)   & 1.0    & 2.2     & 2.5    & 1.2  \\
Final-state radiation (FSR)     & 0.8    & 0.7     & 2.5    & 5.9  \\
ME-PS matching                  & 0.5    & $<$0.1  & 1.8    & 1.9  \\
Underlying event tune (UE)     & 1.5    & 1.5     & 0.4    & 1.4  \\
$\mu_\mathrm{F}/\mu_\mathrm{R}$ scales (ME)              & 0.1    & 0.4     & 0.1    & 1.4  \\
top-\pt                         & 0.2    & 0.4     & 1.6    & 0.3  \\
Ratio $R_{\ttbj/\ttbb}^\mathrm{MC}$ & 1.4    & 0.2     & 1.3     & 0.7  \\[\cmsTabSkip]
\multicolumn{5}{c}{Other uncertainties}\\[\cmsTabSkip]
Pileup        & 0.7    & 0.2 & 1.3    & 0.1  \\
Backgrounds   & 0.3    & 2.0 & 0.7    & 1.2  \\
Simulated sample size & 1.5    & 2.8 & 0.1    & 2.2  \\
Luminosity    & 0.2    & 0.5 & 2.6    & 3.1  \\[\cmsTabSkip]
Total systematic  & 8.0    & 5.5 & 8.8    & 10.0\\
Statistical       &  5.8 &  5.6  &   0.9  & 0.6 \\

\end{tabular}
}  
\label{tab:syst}
\end{table}

\section{Results}
\label{sec:results}
\begin{figure}[htb]
\centering{
\includegraphics[width=0.48\textwidth]{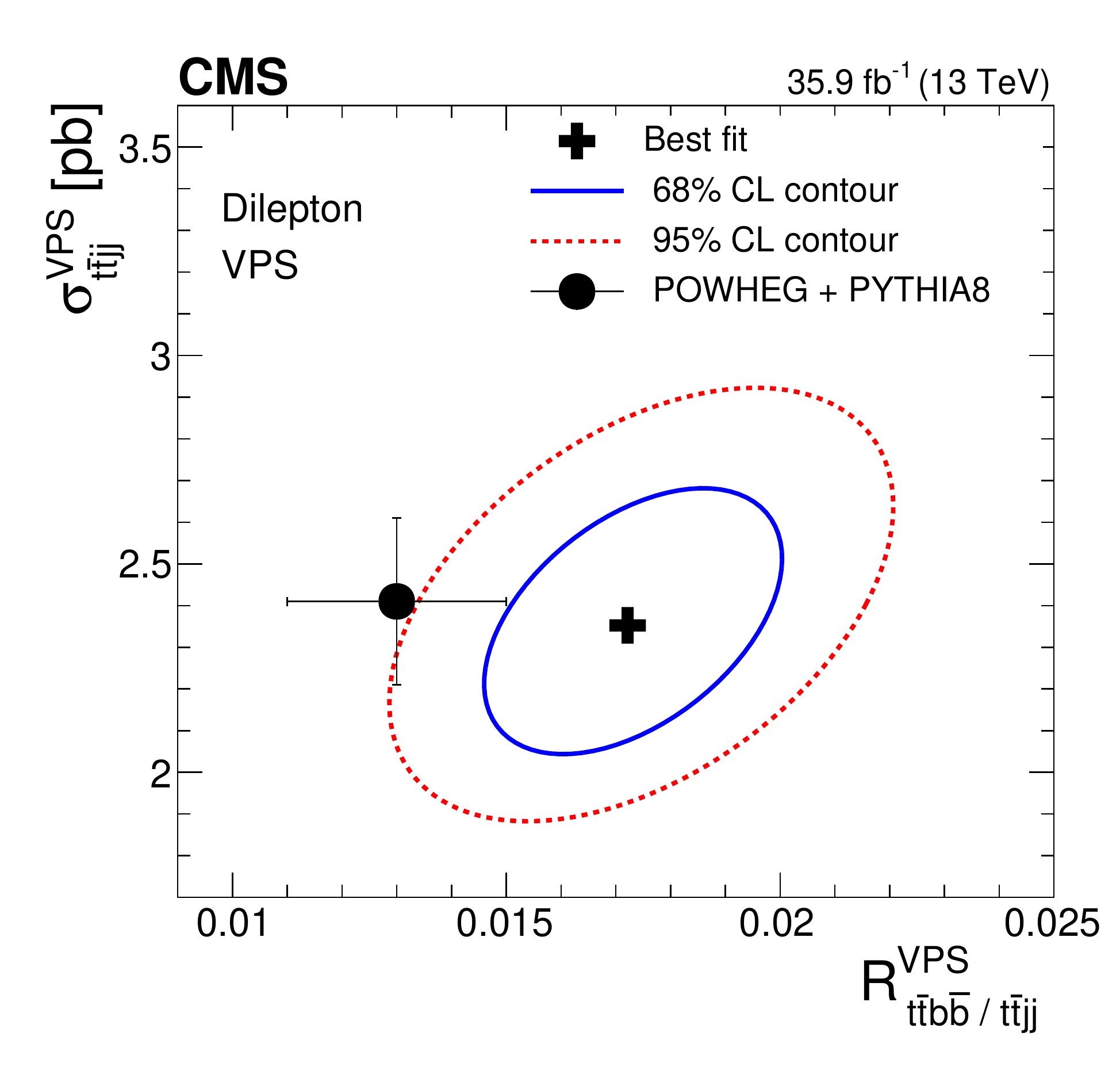}
\includegraphics[width=0.48\textwidth]{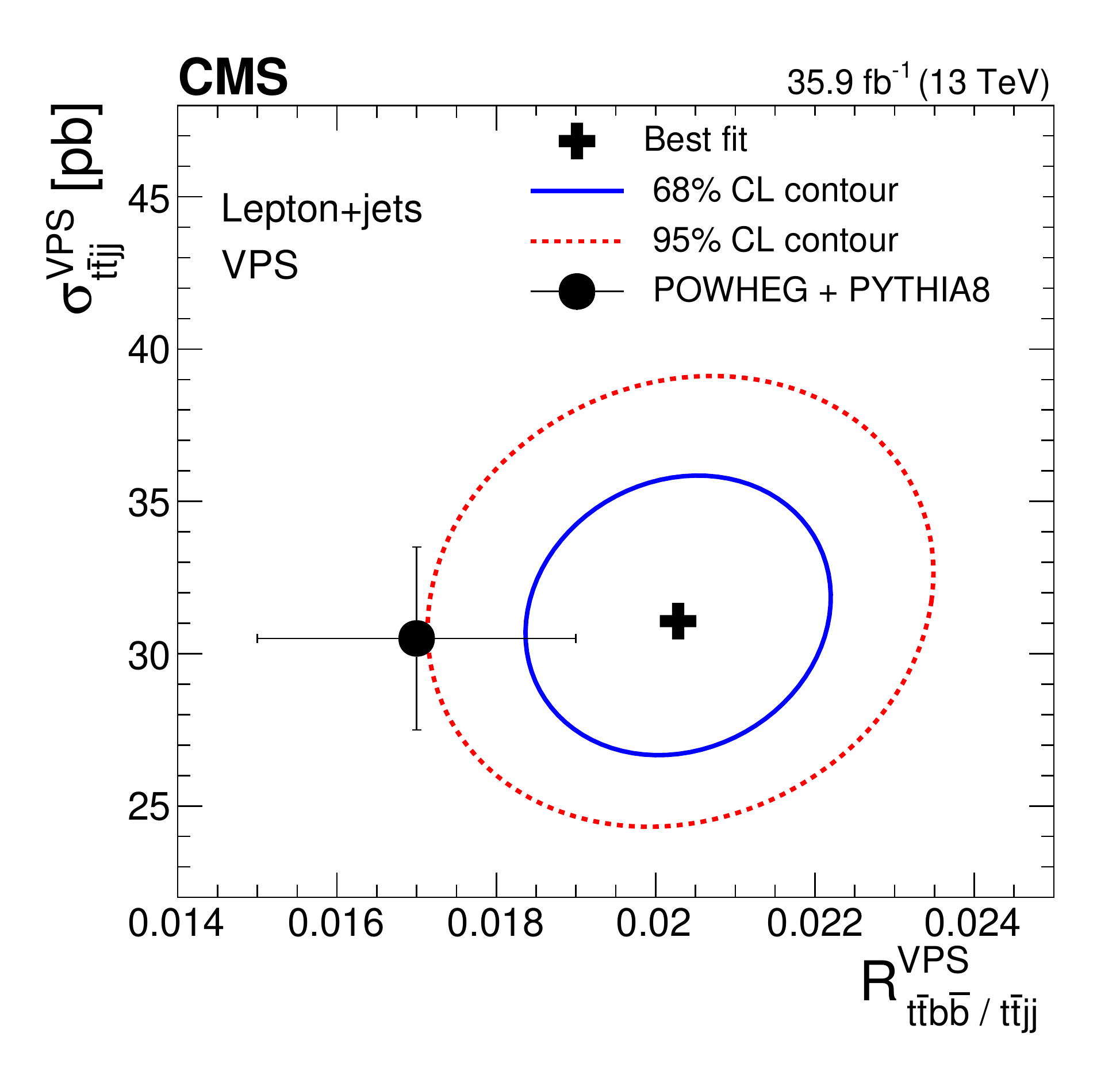}
}
\caption{Results of the simultaneous fit for \rbbjj and \xsecttjj (denoted by the cross) in the visible phase space, along with its 68 and 95\% \CL contours, are shown for the (left) dilepton and (right) lepton+jets channels. 
The solid circle shows the prediction by \POWHEG{} + \PYTHIA{}8.
The uncertainties in the MC prediction are a combination of statistical, $\mu_\mathrm{F}/\mu_\mathrm{R}$ scale, and PDF components; they are assumed to be uncorrelated between $\rbbjj$ and $\xsecttjj$.
} 
\label{fig:2DNLL}
\end{figure}

The simultaneous fit for the $\ttbb$ to $\ttjj$ cross section ratio and the inclusive $\ttjj$ cross section in the VPS for the dilepton (with a jet $\pt>30\GeV$) and lepton+jets (with a jet $\pt>20\GeV$) channels, along with its 68 and 95\% confidence level (\CL) contours are shown in Fig.~\ref{fig:2DNLL}.
The measurements of $\rbbjj$, $\xsecttjj$ and $\xsecttbb$ in the VPS and FPS for both decay channels are given in Table~\ref{tab:result}.
The $\ttbb$ cross section is obtained by multiplying the $\ttjj$ cross section by the $\rbbjj$ cross section ratio.
Table~\ref{tab:result} also includes the expected values from MC simulation and their uncertainties neglecting possible higher-order ME effects.

The uncertainty in the $\ttjj$ cross section is largely dominated by the theoretical uncertainties, such as from the FSR and ME-PS matching, because the final state has a high jet multiplicity. 
The fit sensitivity to high values of the \cPqb tagging discriminant also leads to an important sensitivity to the uncertainties in the \cPqb tagging efficiency, and specifically the light-flavor jet mistagging probability. 
Like the $\ttjj$ cross section, the cross section ratio is also sensitive to the \cPqb tagging efficiency and the variations in the ISR. 
Additionally, both measurements are sensitive to the size of the simulated samples, especially in the lepton+jets channel.  
The two POI, $\xsecttjj$ and $\rbbjj$, show a positive correlation of 48 (12)\% in the dilepton (lepton+jets) channel.
The total uncertainty in the $\ttbb$ cross section for both phase space regions is calculated taking into account these correlations.

Besides the \POWHEG simulation, the measurements of the $\ttjj$ and $\ttbb$ cross sections and their ratio are compared with other MC predictions in Table~\ref{tab:result}.
The \POWHEG predictions for the inclusive $\ttbb$ and $\ttjj$ cross sections in the VPS are in agreement, within the uncertainties, with the measured cross sections in both decay channels.  
The cross section ratio measured in the VPS is larger than the reference \POWHEG prediction by a factor of 1.3 (1.2), with a significance of three (two) standard deviations, in the dilepton (lepton+jets) channel. 
The previous CMS measurement in the dilepton channel at 13\TeV~\cite{078433dfbc4e4b04a600834663f932ea} also reported a larger cross section ratio with respect to the prediction, a factor of 1.8, with a significance of two standard deviations.  

\begin{figure}[hbtp]
\centering{
\includegraphics[width=0.9\textwidth]{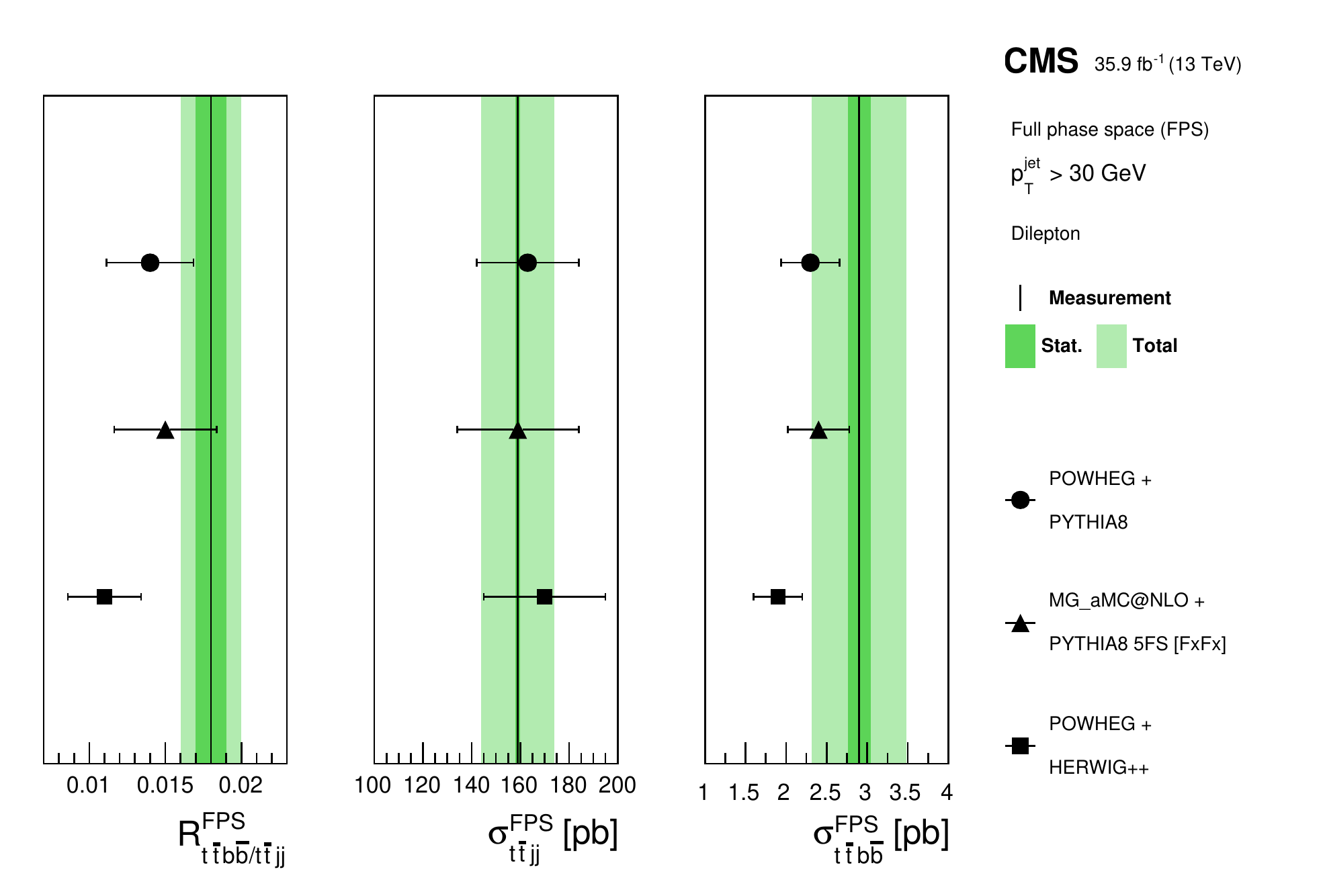}
\includegraphics[width=0.9\textwidth]{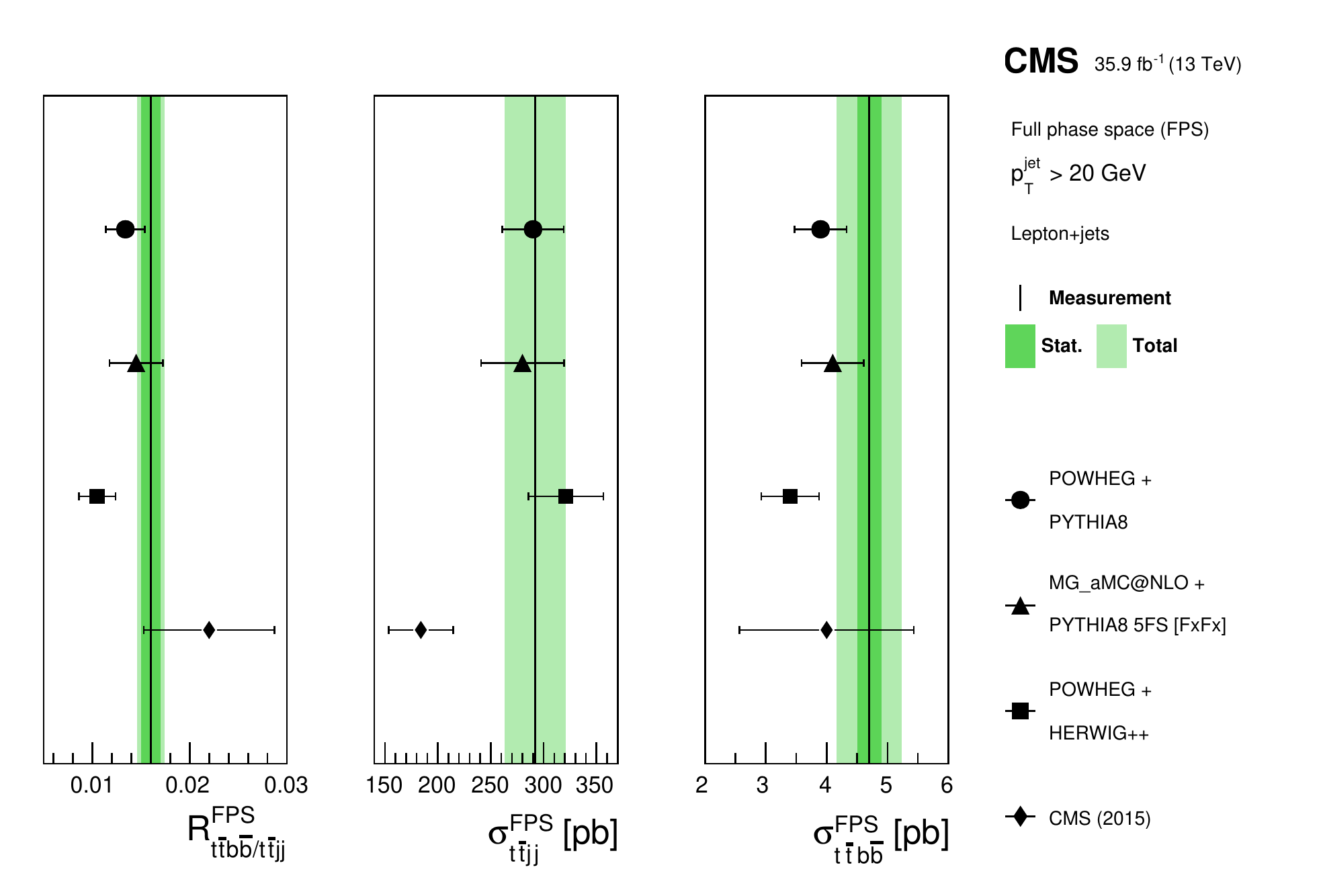}
}
\caption{Measured values (vertical lines) of the $\ttbb$ and $\ttjj$ cross sections and their ratio, along with their statistical and total uncertainties (dark and light bands) in the dilepton (upper) and lepton+jets (lower) channels in the FPS.
Also shown are the theoretical predictions obtained from \POWHEG and \mcNLO (5FS) interfaced with \PYTHIA{}8, and \POWHEG interfaced with \HERWIGpp.
The theoretical predictions for the $\ttjj$ and $\ttbb$ cross sections are normalized to $\sigma_{\ttbar}^\text{\tiny NNLO} = 832\unit{pb}$.  
The previous measurement performed by the CMS Collaboration \cite{078433dfbc4e4b04a600834663f932ea} is also shown with a rhombus marker in the lower plot.
The uncertainties in the MC predictions are a combination of the statistical, $\mu_F/\mu_R$ scale, and PDF components.
}
\label{fig:summ_lj}
\end{figure}

The measurements of the cross sections and their ratio in the FPS is shown in Fig.~\ref{fig:summ_lj}.
Those results are obtained by applying the acceptance correction described in Section~\ref{sec:xsec} to the values measured in the VPS.
The measured inclusive $\ttjj$ and $\ttbb$ cross sections and their ratio for the FPS agree with the MC predictions from \POWHEG and \mcNLO interfaced with \PYTHIA{}8, within the uncertainties, which are larger in the FPS compared to the VPS. 
Predictions from \POWHEG{} + \HERWIG{}++ for the $\ttjj/\ttbb$ cross section ratio, and in consequence for the inclusive $\ttbb$ cross section, are slightly lower than the measured values.
The total relative uncertainties in the $\ttbb$ cross section for the VPS (FPS) are 14 (18)\% for the dilepton channel, and 11 (14)\% for the lepton+jets channel. 
These are the most precise measurements of the inclusive $\ttbb$ and $\ttjj$ cross sections, and their ratio, to date.

\begin{table}[hbtp]
\topcaption{
The measured cross sections $\xsecttbb$ and $\xsecttjj$, and their ratio, for the VPS and FPS, with the results in the latter corrected for the acceptance and branching fractions. 
In both the VPS and FPS definitions, the dilepton and lepton+jets channels require particle-level jets with $\pt>30\GeV$ and 20\GeV, respectively.
The predictions from several MC simulations are also shown. 
The uncertainties in the measurements are split into their statistical (first) and systematic (second) components, while the uncertainties in the MC predictions are a combination of the statistical, $\mu_\mathrm{F}/\mu_\mathrm{R}$ scale, and PDF components.    
}
\label{tab:result}
\centering{
\bgroup
\def\arraystretch{1.2}
\begin{tabular}{lccc}

& $\rbbjj$ & $\xsecttjj$ [\unit{pb}] & $\xsecttbb$ [\unit{pb}]  \\ \hline

\multicolumn{4}{c}{Dilepton channel (VPS)} \\ [\cmsTabSkip]
{{\small\POWHEG{} + \PYTHIA{}8}}  
&$0.013\pm 0.002$ 
&$2.41\pm 0.21$  
&$0.032\pm 0.004$   
\\
Measurement  
&  $0.017\pm 0.001\pm 0.001$ 
&  $2.36\pm 0.02\pm 0.20$
&  $0.040\pm 0.002\pm 0.005$ 
\\[\cmsTabSkip]
\multicolumn{4}{c}{Dilepton channel (FPS)} \\ [\cmsTabSkip]
{\small\POWHEG{} + \PYTHIA{}8}  
&  $0.014\pm 0.003$ 
&  $163\pm 21$ 
&  $2.3\pm 0.4$   
\\
{\small \mcNLO{} + \PYTHIA{8}}  
&  \multirow{2}{*}{$0.015\pm 0.003$} 
&  \multirow{2}{*}{$159\pm 25$}  
&  \multirow{2}{*}{$2.4\pm 0.4$}   
\vspace{-0.2cm}
\\
{\small 5FS [FxFx]} 
& 
&  
&  
\\
{\small\POWHEG{} + \HERWIG{}++}  
&  $0.011\pm 0.002$
&  $170\pm 25$
&  $1.9\pm 0.3$
\\
Measurement    
& $0.018\pm 0.001\pm 0.002$  
& $159\pm 1\pm 15$   
& $2.9\pm 0.1\pm 0.5$   
\\[\cmsTabSkip]
\multicolumn{4}{c}{Lepton+jets channel (VPS)} \\ [\cmsTabSkip]
{\small\POWHEG{} + \PYTHIA{}8}  
&  $0.017\pm 0.002$ 
&  $30.5\pm 3.0$ 
&  $0.52\pm 0.06$   
\\
Measurement  
& $0.020\pm 0.001\pm 0.001$
& $31.0\pm 0.2\pm 2.9$  
& $0.62\pm 0.03\pm 0.07$ 
\\[\cmsTabSkip]
\multicolumn{4}{c}{Lepton+jets channel (FPS)} \\ [\cmsTabSkip]
{\small\POWHEG{} + \PYTHIA{}8}  
&  $0.013\pm 0.002$ 
&  $290\pm 29$ 
&  $3.9\pm 0.4$   
\\
{\small \mcNLO + \PYTHIA{}8}  
&  \multirow{2}{*}{$0.014\pm 0.003$} 
&  \multirow{2}{*}{$280\pm 40$}  
&  \multirow{2}{*}{$4.1\pm 0.4$}   
\vspace{-0.2cm}
\\
{\small 5FS [FxFx]} 
& 
&  
&  
\\
{\small\POWHEG{} + \HERWIG{}++}  
&  $0.011\pm 0.002$ 
&  $321\pm 36$
&  $3.4\pm 0.5$   
\\
Measurement    
& $0.016\pm 0.001\pm 0.001$  
& $292\pm 1\pm 29$  
& $4.7\pm 0.2\pm 0.6$  
\end{tabular}
\egroup
}
\end{table}

\section{Summary} 
\label{sec:summ}
Measurements of the $\ttbb$ and $\ttjj$ cross sections and their ratio are performed independently in the dilepton and lepton+jets final states using a data sample of proton-proton collisions collected at $\sqrt{s} =13\TeV$ by the CMS experiment at the LHC in 2016, and corresponding to an integrated luminosity of $\LumiVal$. 
Leptons and particle-level jets must be in the experimentally accessible kinematic region. 
The inclusive $\ttjj$ cross section and the $\ttbb$ to $\ttjj$ cross section ratio in the fiducial phase space are measured by means of a binned maximum likelihood fit to the \cPqb tagging discriminant distribution of the additional jets, from which the inclusive $\ttbb$ cross section measurement is inferred. 
The cross section ratio and the inclusive $\ttjj$ cross section in the fiducial phase space are extrapolated to the full phase space after correcting for the detector acceptance.

The measured inclusive cross sections in the fiducial phase space for the dilepton and lepton+jets channels, respectively, are $\xsecttbb = 0.040\pm 0.002\stat\pm 0.005\syst\unit{pb}$ and $0.62\pm 0.03\stat\pm 0.07\syst\unit{pb}$, performed by multiplying \xsecttjj with the ratio of \xsecttbb to \xsecttjj, where $\xsecttjj = 2.36\pm 0.02\stat\pm 0.20\syst\unit{pb}$ and $31.0\pm 0.2\stat\pm 2.9\syst\unit{pb}$ and the ratios are $0.017\pm 0.001\stat\pm 0.001\syst$ and $0.020\pm 0.001\stat\pm 0.001\syst$. 
The treatment of the systematic uncertainties as nuisance parameters in the fit leads to an improvement in the precision compared to previous measurements. 
The inclusive $\ttbb$ cross sections and the cross section ratios for both decay channels measured in the full phase space have values higher than, but consistent with, the predictions from several different Monte Carlo generators.
A measured $\ttbb$ cross section higher than Monte Carlo predictions is also reported in a recent measurement performed by the CMS Collaboration in the fully hadronic final state~\cite{CMS-PAS-TOP-18-011}.  

\begin{acknowledgments}

    We congratulate our colleagues in the CERN accelerator departments for the excellent performance of the LHC and thank the technical and administrative staffs at CERN and at other CMS institutes for their contributions to the success of the CMS effort. In addition, we gratefully acknowledge the computing centers and personnel of the Worldwide LHC Computing Grid for delivering so effectively the computing infrastructure essential to our analyses. Finally, we acknowledge the enduring support for the construction and operation of the LHC and the CMS detector provided by the following funding agencies: BMBWF and FWF (Austria); FNRS and FWO (Belgium); CNPq, CAPES, FAPERJ, FAPERGS, and FAPESP (Brazil); MES (Bulgaria); CERN; CAS, MoST, and NSFC (China); COLCIENCIAS (Colombia); MSES and CSF (Croatia); RPF (Cyprus); SENESCYT (Ecuador); MoER, ERC IUT, PUT and ERDF (Estonia); Academy of Finland, MEC, and HIP (Finland); CEA and CNRS/IN2P3 (France); BMBF, DFG, and HGF (Germany); GSRT (Greece); NKFIA (Hungary); DAE and DST (India); IPM (Iran); SFI (Ireland); INFN (Italy); MSIP and NRF (Republic of Korea); MES (Latvia); LAS (Lithuania); MOE and UM (Malaysia); BUAP, CINVESTAV, CONACYT, LNS, SEP, and UASLP-FAI (Mexico); MOS (Montenegro); MBIE (New Zealand); PAEC (Pakistan); MSHE and NSC (Poland); FCT (Portugal); JINR (Dubna); MON, RosAtom, RAS, RFBR, and NRC KI (Russia); MESTD (Serbia); SEIDI, CPAN, PCTI, and FEDER (Spain); MOSTR (Sri Lanka); Swiss Funding Agencies (Switzerland); MST (Taipei); ThEPCenter, IPST, STAR, and NSTDA (Thailand); TUBITAK and TAEK (Turkey); NASU (Ukraine); STFC (United Kingdom); DOE and NSF (USA). 
    
    \hyphenation{Rachada-pisek} Individuals have received support from the Marie-Curie program and the European Research Council and Horizon 2020 Grant, contract Nos.\ 675440, 752730, and 765710 (European Union); the Leventis Foundation; the A.P.\ Sloan Foundation; the Alexander von Humboldt Foundation; the Belgian Federal Science Policy Office; the Fonds pour la Formation \`a la Recherche dans l'Industrie et dans l'Agriculture (FRIA-Belgium); the Agentschap voor Innovatie door Wetenschap en Technologie (IWT-Belgium); the F.R.S.-FNRS and FWO (Belgium) under the ``Excellence of Science -- EOS" -- be.h project n.\ 30820817; the Beijing Municipal Science \& Technology Commission, No. Z191100007219010; the Ministry of Education, Youth and Sports (MEYS) of the Czech Republic; the Deutsche Forschungsgemeinschaft (DFG) under Germany's Excellence Strategy -- EXC 2121 ``Quantum Universe" -- 390833306; the Lend\"ulet (``Momentum") Program and the J\'anos Bolyai Research Scholarship of the Hungarian Academy of Sciences, the New National Excellence Program \'UNKP, the NKFIA research grants 123842, 123959, 124845, 124850, 125105, 128713, 128786, and 129058 (Hungary); the Council of Science and Industrial Research, India; the HOMING PLUS program of the Foundation for Polish Science, cofinanced from European Union, Regional Development Fund, the Mobility Plus program of the Ministry of Science and Higher Education, the National Science Center (Poland), contracts Harmonia 2014/14/M/ST2/00428, Opus 2014/13/B/ST2/02543, 2014/15/B/ST2/03998, and 2015/19/B/ST2/02861, Sonata-bis 2012/07/E/ST2/01406; the National Priorities Research Program by Qatar National Research Fund; the Ministry of Science and Education, grant no. 14.W03.31.0026 (Russia); the Tomsk Polytechnic University Competitiveness Enhancement Program and ``Nauka" Project FSWW-2020-0008 (Russia); the Programa Estatal de Fomento de la Investigaci{\'o}n Cient{\'i}fica y T{\'e}cnica de Excelencia Mar\'{\i}a de Maeztu, grant MDM-2015-0509 and the Programa Severo Ochoa del Principado de Asturias; the Thalis and Aristeia programs cofinanced by EU-ESF and the Greek NSRF; the Rachadapisek Sompot Fund for Postdoctoral Fellowship, Chulalongkorn University and the Chulalongkorn Academic into Its 2nd Century Project Advancement Project (Thailand); the Kavli Foundation; the Nvidia Corporation; the SuperMicro Corporation; the Welch Foundation, contract C-1845; and the Weston Havens Foundation (USA). 
\end{acknowledgments}

\bibliography{auto_generated}
\cleardoublepage \appendix\section{The CMS Collaboration \label{app:collab}}\begin{sloppypar}\hyphenpenalty=5000\widowpenalty=500\clubpenalty=5000\vskip\cmsinstskip
\textbf{Yerevan Physics Institute, Yerevan, Armenia}\\*[0pt]
A.M.~Sirunyan$^{\textrm{\dag}}$, A.~Tumasyan
\vskip\cmsinstskip
\textbf{Institut f\"{u}r Hochenergiephysik, Wien, Austria}\\*[0pt]
W.~Adam, F.~Ambrogi, T.~Bergauer, M.~Dragicevic, J.~Er\"{o}, A.~Escalante~Del~Valle, M.~Flechl, R.~Fr\"{u}hwirth\cmsAuthorMark{1}, M.~Jeitler\cmsAuthorMark{1}, N.~Krammer, I.~Kr\"{a}tschmer, D.~Liko, T.~Madlener, I.~Mikulec, N.~Rad, J.~Schieck\cmsAuthorMark{1}, R.~Sch\"{o}fbeck, M.~Spanring, D.~Spitzbart, W.~Waltenberger, C.-E.~Wulz\cmsAuthorMark{1}, M.~Zarucki
% ...
\vskip\cmsinstskip
\dag: Deceased\\
1:  Also at Vienna University of Technology, Vienna, Austria\\
% ...
\end{sloppypar}
\end{document}